\documentclass[aps,prb,twocolumn,showpacs,superscriptaddress]{revtex4-2}

\usepackage{amssymb,amsmath,amsthm}
\usepackage{mathtools}
\usepackage{esint}
\usepackage{siunitx}
\usepackage{yfonts}

\usepackage{graphicx}
\usepackage{multirow}

\usepackage{xcolor}

\usepackage{hyperref}

\usepackage[nameinlink]{cleveref}

\newtheorem{theorem}{Theorem}
\graphicspath{{./IMG/}}
\newcommand{\occf}{f}

\begin{document}

\title{Low-temperature behavior of density-functional theory for metals based on density-functional perturbation theory and Sommerfeld expansion}

\author{Xavier Gonze*}
\affiliation{European Theoretical Spectroscopy Facility, Institute of Condensed Matter and Nanosciences, Universit\'{e} catholique de Louvain, Chemin des \'{e}toiles 8, bte L07.03.01, B-1348 Louvain-la-Neuve, Belgium.}

\author{Christian Tantardini*}
\affiliation{Center for Integrative Petroleum Research, King Fahd University of Petroleum and Minerals, Dhahran 31261, Saudi Arabia}

\author{Antoine Levitt*}
\affiliation{Laboratoire de math\'ematiques d'Orsay, Universit\'{e} Paris-Saclay, France.}

\email{xavier.gonze@uclouvain.be, 
\\ christiantantardini@ymail.com,
\\ antoine.levitt@universite-paris-saclay.fr}

\date{\today}

\begin{abstract}
The temperature dependence of most solid-state properties is dominated by lattice vibrations, but metals display notable purely electronic effects at low temperature, such as the linear specific heat and the linear entropy, that were derived by Sommerfeld for the non-interacting electron gas via the 
low-temperature expansion of Fermi–Dirac integrals. 
Here we  treat temperature as a perturbation within density-functional perturbation theory (DFPT). 
For finite temperature,
we show how self-consistency screens the bare, temperature-induced density change obtained in the non-interacting picture: the inverse transpose of the electronic dielectric operator, that includes Adler-Wiser and a term related to the shift in Fermi level, links the self-consistent density response to the bare thermal density change.
This approach is implemented in DFTK, and demonstrated by the computation of the second-order derivative of the free energy, and the first-order derivative of entropy for aluminum.
Then, we examine the $T\!\to\!0$ limit. The finite temperature formalism contains divergences, that we cure using the Sommerfeld expansion to analyze metallic systems at 0 K. 
The electronic free energy is quadratic in $T$ provided the Fermi level is not at a Van Hove singularity of the density of states. If the latter happens, another temperature behavior might appear, depending on the type of Van Hove singularity, that we analyze. 
 Our formulation applies to systems periodic in one, two, or three dimensions, and provides a basis for studying temperature-dependent electronic instabilities (e.g., charge-density waves) within density-functional theory and DFPT.
\end{abstract}

\maketitle

\section{Introduction}

Nearly one century ago, Sommerfeld presented a simple model for metals~\cite{Sommerfeld1928} that was able to account for their low-temperature linear-$T$ specific heat - a purely electronic effect.
This textbook result~\cite{ashcroft1976} is at variance with most temperature-dependent properties of materials, for which phonon effects dominate, at least at room temperature. The latters can be derived from Bose-Einstein statistics, and yield e.g.
the cubic-$T$ specific heat of insulators.

In order to prove such linear-$T$ behavior, Sommerfeld considered energy integrals whose arguments are product of continuous, temperature-independent functions times the temperature-dependent Fermi-Dirac function. The latter 
is discontinuous at $T$=0, but Sommerfeld  provided the relevant low-$T$ expansion of such energy integrals.
On the basis of this mathematical result, the linear temperature dependence of the specific heat and 
entropy for the homogeneous electron gas was derived, as well as the temperature dependence of the free energy, a quadratic departure from the zero-temperature value. 
Generalizing such behavior beyond Sommerfeld's homogeneous electron gas model is obvious, provided one works in the non-interacting electron approximation, and provided that the density of state at the Fermi level is finite, continuous and derivable.

However, such behavior might be modified by the electron-electron interactions, and also if the density of states has singularities at the Fermi energy. 
For example, it is well-known that the Hartree-Fock description of metals is pathological, with a spurious vanishing density of states at the Fermi level,~\cite{inkson1986} and indeed the corresponding specific heat is sublinear and not linear. The situation in Density-Functional Theory 
(DFT),\cite{Martin2004} aligns with the non-interacting case, at least for the usual approximations, like local density approximation (LDA) and generalized-gradient approximation (GGA), for which the density of states at the Fermi level does not spuriously vanish. 

The treatment of finite temperature in DFT is done routinely in publicly available software applications, although 
sometimes with some additional difficulties compared to the treatment of gapped systems
\cite{DosSantos2023}.
Examining the variation of properties with temperature is however done using finite differences presently. 
Still, there is a natural framework to treat generic variations around some reference situation in DFT, namely
Density-Functional Perturbation Theory (DFPT)\cite{Baroni1987,Gonze1997,Gonze1997a,Baroni2001,Gonze2005a}.
DFPT is a workhorse for the treatment of adiabatic perturbations (e.g. phonons, electronic fields, magnetic fields, strains) in crystalline materials, nanosystems and molecules.
It is available in several widely used first-principles codes\cite{Giannozzi2017,Gonze_2020,Romero2020}. 
While the basic concepts of DPFT for metals had been proposed a few
decades ago\cite{Gironcoli1995}, it is only quite recently that the
variational formulation of DFPT for metals has been exposed in
detail\cite{Gonze2024a} and the impact of a gauge choice assessed
critically\cite{cances2023numerical}. Also, the formulation and application of the 2n+1 theorem of DFPT\cite{Gonze1989_2nplus1,Gonze1995_PRA1086,Gonze1995_PRA1096} for metals has only been recently published\cite{Zabalo2024a}.

In this work, we present the treatment of a change of temperature within DFPT. 
A bare change of density caused by the temperature change replaces the change in external potential (or applied field) in term of which DFPT is usually formulated. Such bare change of density must then be screened self-consistently. 
This treatment is quite easy when the temperature
variation happens around a finite temperature, but much less when considering
a departure from strictly zero temperature.

The bare modification of the density due to the change of temperature, be it at finite temperature or at zero Kelvin,
induces then a change of potential, that itself induces a density response. The latter is not only based on the well-known Adler-Wiser independent-particle susceptibility\cite{Adler1962,Wiser1963}, but includes a response at the Fermi level.
Such contribution had already been noted in the analysis of the
convergence of DFT calculations\cite{Herbst2020a}.

We then analyze the 
low-temperature behavior of metals
by incorporating the Sommerfeld expansion in the bare change of density and in the self-consistency procedure. 
If the $T=0$ chemical potential (or Fermi energy) is not precisely at a Van Hove singularity of the DOS\cite{VanHove1953}, the temperature dependence of the bare change of density is quadratic in the temperature, and the screening does not modify such behavior.
By contrast, if the $T=0$ chemical potential (or Fermi energy) is precisely at a Van Hove singularity, the temperature dependence of the bare change of density has another behavior, depending on the type of Van Hove singularity. 

Taking the Sommerfeld expansion into account is essential for understanding the temperature dependence of DFT properties of metals at low temperatures.
We foresee that this approach might be applied for the temperature-dependent study 
of charge-density waves, or other purely electronic phenomena, 
taking into account the electron-electron interaction at the mean-field level. 
The temperature-dependence of properties usually obtained from DFPT for metals~\cite{Gironcoli1995,Gonze2024a} might also be treated subsequently, for example using the 2n+1 theorem with several perturbations, including the temperature perturbation.

The paper is organized as follows. Sec. \ref{Sec:VDFTM} reviews the variational formulation of density-functional theory (DFT) for metals.
A temperature change applied when the reference temperature does not vanish is treated using density-functional perturbation theory in Sec. \ref{sec:DFPT}.
Validation of this DFPT formalism is provided in Sec. \ref{sec:Validation}.
Sec. \ref{sec:Sommerfeld} introduces the Sommerfeld expansion in the low-temperature limit, first for the free-electron gas and then taking into account the self-consistency. The effect of Van Hove singularities on the Sommerfeld expansion is also treated.
Sec. \ref{sec:lowT-DFT} builds on these results to develop the low-temperature DFT for metals, in the limit of zero electronic temperature.

The supplemental material includes seven sections: a discussion of temperature-dependent exchange-correlation functionals (SuppMat1); the development of Density-Functional Perturbation Theory of a temperature change for finite systems (SuppMat2);
mathematical information about the Sommerfeld expansion (SuppMat3);
the proof of theorem 1 about the convergence of the Sommerfeld series (SuppMat4);
the combination of Sommerfeld approach with perturbation theory (SuppMat5);
the combination of Sommerfeld approach with density-functional perturbation theory (SuppMat6);
the illustration of the effect of a Van Hove singularity on the temperature dependence of the free energy (SuppMat7).
\\
\\



\section{Density-functional theory for metallic periodic solids}
\label{Sec:VDFTM}

In this section, the notations and conventions for metallic periodic solids are introduced. 
They are similar to 
those in Sec.V of Ref. \onlinecite{Gonze2024a}.
We also recall the variational formulation of DFT for metallic periodic solids, as introduced in 1997 by Marzari, Vanderbilt and Payne (MVP)\cite{Marzari1997}.

The MVP electronic free (Helmholtz) energy per unit cell writes
\begin{widetext}
\begin{eqnarray}
F_\textrm{el}[T;
\{u_{n
\textbf{k}}
\},\{\rho_{nm
\textbf{k}}
\}
]
&=& 
n_{\textrm{s}} \Omega_0
\int_{\textrm{BZ}}
\sum_{nm}\rho_{nm\textbf{k}} \langle u_{m\textbf{k}} | \hat{K}_{\textbf{k}\textbf{k}} + \hat{v}_{\textrm{ext},\textbf{k}\textbf{k}} | u_{n\textbf{k}} \rangle
[d\textbf{k}]
+
E_{\textrm{Hxc}}[\rho] 
- TS\big(\{\rho_{nm\textbf{k}}\}\big).
\label{eq:F_metal}
\end{eqnarray}
\end{widetext}

In the MVP expression, the sums over the band indices $n$ and $m$ extend to infinity, $\hat{K}$ is the kinetic energy operator, $\hat{v}_{\textrm{ext}}$ is the external
potential (e.g. created by the nuclei, as well as any other  additional external potential), $E_{\textrm{Hxc}}$ is the DFT Hartree and exchange-correlation (XC) energy functional of the density $\rho(\textbf{r})$, per unit cell, $T$ the temperature, and $S$ the electronic entropy per unit cell. 
Operators are denoted with a circumflex accent, and, later, the real-space kernel of an operator $\hat{A}$ will be written $A(\textbf{r},\textbf{r}')$. 
$\textbf{k}$ labels wavevectors in the Brillouin Zone (BZ),
 $[d\textbf{k}]=d\textbf{k}/(2\pi)^D$ and $D$ is the dimensionality of the system.
For sake of simplicity, spin-unpolarized systems are considered, where $n_{\textrm{s}}$=2 accounts for the spin degeneracy.

In principle, the exchange-correlation energy should be temperature dependent, giving the Mermin functional \cite{mermin1965thermal, Parr1989, Gonis2018}. 
This temperature dependence
is however not the focus of this paper, at variance with the
direct impact of the temperature multiplying the entropy in the last term of Eq.~(\ref{eq:F_metal}).
In the section SuppMat1, we briefly discuss the temperature dependence of some existing XC functionals~\cite{Perrot1984, Brown2013, Karasiev2014, Karasiev2018}, and its consequence on the present work.
\\

The MVP free energy, Eq.~(\ref{eq:F_metal}), is to be minimized with respect to the
periodic part of the (trial) Bloch wavefunctions, $u_{n\textbf{k}}$, as well as
with respect to the
(trial) matrix representation 
$\{\rho_{nm\textbf{k}}\}$ 
of the one-particle density matrix operator in this set of (trial) wavefunctions.

The wavefunctions are normalized as follows,
\begin{equation}
\langle u_{m\textbf{k}} | u_{n\textbf{k}} \rangle=
\frac{1}{\Omega_0}
\int_{\Omega_0}
u_{m\textbf{k}}(\textbf{r})^*
u_{n\textbf{k}}(\textbf{r})
d\textbf{r}=\delta_{mn},
\label{eq:ortho}
\end{equation}
where $\Omega_0$ is the volume of the primitive periodic cell in three dimensions (or surface in two dimensions, or length in one dimension).
The volume of the Brillouin Zone is 
$\Omega_\textrm{BZ}=(2\pi)^D / \Omega_0$.

The matrix elements of the kinetic operator and external potential operator are evaluated over the primitive cell. 
The expression of the electronic density relies on the
one-particle density matrix elements,
\begin{eqnarray}
\rho(\textbf{r}) = 
n_{\textrm{s}}
\int_{\textrm{BZ}}
\sum_{nm}\rho_{nm\textbf{k}}
u_{m\textbf{k}}^*(\textbf{r})
u_{n\textbf{k}}(\textbf{r})
[d\textbf{k}].
\label{eq:n_metal_densitymatrix}
\end{eqnarray}
This electronic density is periodic.
The number of electrons per unit cell
is
\begin{eqnarray}
N_e=\int_{\Omega_0}\rho(\textbf{r}) d\textbf{r}.
\label{eq:N_e}
\end{eqnarray}

The entropy per unit cell $S$ in Eq.(\ref{eq:F_metal}) is expressed in terms of the
eigenvalues $f_{n\textbf{k}}$,
of the density matrix
\begin{equation}
\sum_{m'}\rho_{mm'\textbf{k}}
\mathfrak{f}_{m'n\textbf{k}}=
f_{n\textbf{k}}
\mathfrak{f}_{mn\textbf{k}},
\end{equation}
where $\mathfrak{f}_{m'n\textbf{k}}$ are the components of the corresponding eigenvectors.

As follows,
\begin{equation}
S\big(\{\rho_{nm\textbf{k}}\}\big)=
n_{\textrm{s}}\Omega_0
\int_{\textrm{BZ}}
\sum_{n}
k_\textrm{B}s_\textrm{FD}\big(f_{n\textbf{k}}\big)
[d\textbf{k}],
\label{eq:entropy_4}
\end{equation} 
where the single-orbital Fermi-Dirac (FD) entropy is
\begin{equation}
s_\textrm{FD}(f) = -\Big(f \ln(f) + (1-f) \ln(1-f) \Big),
\label{eq:sFD}
\end{equation}
and $k_\textrm{B}$ is Boltzmann's constant.

Minimizing the free energy while enforcing
the condition of constant number of electrons per unit cell, $N_e$, Eq.(\ref{eq:N_e}) and orthonormalization constraints, Eq.(\ref{eq:ortho}),
can be done
using the Lagrange multiplier method, 
introducing the multiplier $\mu$ (identified to the chemical potential) 
to preserve $N_e$, and the array of multipliers $\Lambda_{nm\textbf{k}}$
to preserve the orthonormalization constraints.
The expression of the free energy, 
augmented with the Lagrange multiplier terms, denoted $F^+$, 
is presented in Ref.\onlinecite{Gonze2024a}, see Eq.(16) and (20), as well as Eq.(S4) of the section SuppMat2.
As shown by MVP, at the minimum, one has
the following equation,
\begin{eqnarray}
\hat{H}_{\textbf{k}\textbf{k}} | u_{n\textbf{k}} \rangle &=&
\big( \hat{K}_{\textbf{k}\textbf{k}} + \hat{v}_{\textrm{ext},\textbf{k}\textbf{k}} 
+ \hat{v}_{\textrm{Hxc}}[\rho] 
\big)
| u_{n\textbf{k}} \rangle
\nonumber\\
&=& \sum_m \Lambda_{nm\textbf{k}} | u_{m\textbf{k}} \rangle,
\label{eq:KohnSham_metal_Lagrange}
\end{eqnarray}
where 
$\hat{v}_{\textrm{Hxc}}[\rho]$
is the local potential operator obtained from the
functional derivative of 
$E_{\textrm{Hxc}}[\rho]$ with respect to
$\rho(\textbf{r})$, and $\hat{H}_{\textbf{k}\textbf{k}}$ is the Hamiltonian. It is periodic.
Also, at the minimum, the Hamiltonian and density matrix
commute, and can be simultaneously diagonalized, as shown by MVP.
The set of $| u_{m\textbf{k}} \rangle$ is indeed
chosen, thanks to a unitary transformation (gauge freedom), to satisfy the
Kohn-Sham equations
\begin{eqnarray}
\hat{H}_{\textbf{k}\textbf{k}} | u_{n\textbf{k}} \rangle= \varepsilon_{n\textbf{k}} | u_{n\textbf{k}} \rangle.
\label{eq:KohnSham_metal}
\end{eqnarray}
We continue to focus on the situation at the minimum of the free energy, and, in order to emphasize the role of temperature, we indicate now, and until the end of the section, the direct or indirect temperature-dependence of the different quantities.

The relationship between eigenenergies and occupation numbers is obtained, namely,
\begin{eqnarray}
\occf_{n\textbf{k}}(T,\mu(T)) &=& f_{\textrm{FD}}
\big( (\varepsilon_{n\textbf{k}}(T)-\mu(T))
    /k_\textrm{B}T \big),
\label{eq:fnk_FD}
\end{eqnarray}
with
\begin{eqnarray}
f_{\textrm{FD}}(x)&=& \big( \exp(x)+1 \big)^{-1}.
\label{eq:FD}
\end{eqnarray}
It decreases monotonically from 1 to 0. Ref.\onlinecite{Gonze2024a} used another definition, with $f_{\rm FD}(x)$ increasing
monotonically from 0 to 1, its argument having changed sign.

The density can be expressed in terms of the usual occupation numbers, 
\begin{align}
\rho(T,\textbf{r}) =& 
n_{\textrm{s}}
\int_{\textrm{BZ}}
\sum_{n}
\nonumber\\
&\occf_{n\textbf{k}}(T,\mu(T))
u_{n\textbf{k}}^*(T,\textbf{r})
u_{n\textbf{k}}(T,\textbf{r})
[d\textbf{k}].
\label{eq:n_metal_diagonal}
\end{align}

For later use in the $T \to 0$ limit, we reformulate the Brillouin Zone integral entering the electronic density as an energy integral.
The energy-resolved electronic density (periodic) is defined as
\begin{eqnarray}
\rho(T,\varepsilon,\textbf{r}) = 
n_{\textrm{s}}
\int_{\textrm{BZ}}
\sum_{n}
\delta(\varepsilon-\varepsilon_{n\textbf{k}}(T))
\rho_{n\textbf{k}}(T,\textbf{r})
[d\textbf{k}],
\label{eq:n_e_metal_1}
\end{eqnarray}
where the state-electronic density is 
\begin{eqnarray}
\rho_{n\textbf{k}}(T,\textbf{r})=
u_{n\textbf{k}}^*(T,\textbf{r})
u_{n\textbf{k}}(T,\textbf{r}).
\label{eq:rho_nk_1}
\end{eqnarray}
The latter is a real periodic function of $\textbf{r}$, with
\begin{eqnarray}
\Omega_0= 
\int_{\Omega_0} \rho_{n\textbf{k}}(T,\textbf{r})
d\textbf{r}, 
\label{eq:int_rhonk_1}
\end{eqnarray}
due to Eq.(\ref{eq:ortho}).
It should not be mistaken for the density matrix 
$\rho_{nm\textbf{k}}$, despite a similar notation (except that the number of indices differs and the argument $(T,\textbf{r})$ is present for the former). 
We will no longer encounter $\rho_{nm\textbf{k}}$
in the next sections,
so the risk of confusion is limited.

Integrating over the energy, one gets
\begin{eqnarray}
\rho(T,\textbf{r}) = 
\int_{-\infty}^{+\infty}
f_{\textrm{FD}}\bigg(\frac{\varepsilon - \mu(T)}{k_\textrm{B}T} \bigg)
\rho(T,\varepsilon,\textbf{r})
d\varepsilon . 
\label{eq:n_metal_int_e_1}
\end{eqnarray}

The Density-Of-States (DOS) is a related quantity,
\begin{align}
g_{\textrm{DOS}}(T,\varepsilon)=& 
n_{\textrm{s}}\Omega_0
\int_{\textrm{BZ}}
\sum_{n}
\delta(\varepsilon-\varepsilon_{n\textbf{k}}(T))
[d\textbf{k}]
\\
=&\int_{\Omega_0} \rho(T,\varepsilon,\textbf{r})
d\textbf{r} . 
\label{eq:gDOS_e_inhom_1}
\end{align}
Thus, the quantity $\rho(T,\varepsilon,\textbf{r})$ can also be referred to as a local density of states (l-DOS), since it delivers the DOS for a specific point in space, and its spatial integral gives the DOS. 


\section{DFPT : changing the temperature at finite temperature}
\label{sec:DFPT}

We now consider the expansion of the density, Hamiltonian,
eigenenergies, around a reference temperature
$T^{(0)}$, as a function of a temperature change $\Delta T=T-T^{(0)}$ around this reference temperature.

We write generically, up to second order,
\begin{align}
X(\Delta T) = X^{(0)}+(\Delta T) X^{(1)}
+
(\Delta T)^2 X^{(2)}
+
\mathcal{O}\big( (\Delta T)^{3}\big).
\nonumber\\
\label{eq:T0_Xgeneric}
\end{align}

The section SuppMat2 presents the case of finite systems,
and points out the differences between the treatment of a temperature change and the one of an external potential change, the latter having been the focus of the variational formulation of DFPT for metals in Ref.\onlinecite{Gonze2024a}.
In this main text, instead, we focus on periodic metals. We use the
diagonal gauge\cite{Gonze2024a}, i.e. we impose that the perturbed orbitals diagonalize the perturbed Hamiltonian. 
The diagonal gauge is numerically unstable and assumes
non-degeneracy of the eigenvalues; however,  we use it for simplicity
since the end formulas that are
the main result of this paper are not sensitive to the choice of
gauge.
Also, compared to the DFPT theory for metals of Ref.\onlinecite{Gonze2024a}, in which generic perturbations, possibly non-commensurate with the periodicity of the crystal have been considered, a perturbative change of temperature does not change the crystalline periodicity, provided one is away from electronic, magnetic or orbital phase
transition temperature.

The first-order change of electronic free energy per unit cell, due to a temperature change is given by
\begin{align}
\frac{\partial F_\textrm{el}}{\partial T}_{|T=T^{(0)}}
= &
F_\textrm{el}^{+(1)} [T^{(0)}]  
= 
-S\big(\{f_{n\textbf{k}}^{(0)}\delta_{mn}\}\big)
\label{eq:F1=-S}\\
=&
-
n_{\textrm{s}}\Omega_0
\int_\textrm{BZ}
\sum_n
k_\textrm{B}s_\textrm{FD}\big(f_{n\textbf{k}}^{(0)}\big)
[d\textbf{k}].\label{eq:F1el}
\end{align}
$F_{\textrm{el}}^{+}$ is the free energy, augmented with Lagrange multipliers, see Eq.(S4) of the section SuppMat2.
This result is a well-known thermodynamic identity 
$\partial F_\textrm{el} / \partial T=-S$
and also aligns with the 
Hellmann-Feynman theorem\cite{Hellmann1937,Feynman1939}, in that
the first-order changes of 
wavefunctions or occupation numbers are not needed
in order to compute the first derivative of the 
variational free energy with respect to a perturbation.

The second-order change of electronic free energy over unit cell, due to a temperature change, is given by the following variational expression:
\begin{widetext}
\begin{align}
F_\textrm{el}^{+(2)} [T^{(0)}, \{ u_{\textrm{d}m\textbf{k}}^{(1)} \}, \{ f_{m\textbf{k}}^{(1)} \} ]  
= & n_\textrm{s} \Omega_{0} \int_{\textrm{BZ}} 
\sum_{m}\Bigg(
f_{m\textbf{k}}^{(0)} \langle u_{\textrm{d}m\textbf{k}}^{(1)} | \hat{H}_{\textbf{k}\textbf{k}}^{(0)} 
- \varepsilon_{m\textbf{k}}^{(0)}
| u_{\textrm{d}m\textbf{k}}^{(1)} \rangle
 \nonumber
 \\ 
 & 
\quad \quad \quad \quad \quad \quad \quad - \frac{1}{2} \frac{\partial \varepsilon}{\partial f} \Bigg|_{f_{m\textbf{k}}^{(0)}} \big(
f_{m\textbf{k}}^{(1)}
\big)^{2}   
-\Big( \frac{\varepsilon_{m\textbf{k}}^{(0)}-\mu^{(0)}}{T^{(0)}} +\mu^{(1)} \Big) f_{m\textbf{k}}^{(1)}
 \Bigg) [d\textbf{k}] 
 \nonumber \\
& + \frac{1}{2} \int_{\Omega_{0}} \int  K_\textrm{Hxc}(\textbf{r},\textbf{r'}) \rho^{*(1)}(\textbf{r}) \rho^{(1)}(\textbf{r'}) d\textbf{r} d\textbf{r'}. \label{eq:177}
\end{align}
\end{widetext}
In Eq.~\eqref{eq:177}, 
$u_{\textrm{d}m\textbf{k}}^{(1)}$
is the change of the periodic part of the Bloch wavefunctions for band $m$ and wavevector 
$\textbf{k}$, in the diagonal gauge, while
$f_{m\textbf{k}}^{(1)}$
denotes the change of the corresponding diagonal element of the density matrix. 
The latter is also equal to the change of occupation number, the quantity relevant in the diagonal gauge (off-diagonal elements of the density matrix are zero anyway in this gauge). This explains the choice of notation. 
$F_\textrm{el}^{+(2)} [T^{(0)}, \{ u_{\textrm{d}m\textbf{k}}^{(1)} \}, \{ f_{m\textbf{k}}^{(1)} \} ]$ obviously also depends on the
zero-order wavefunctions and occupation numbers,
but this dependence is not explicitly mentioned
in its arguments, a common practice in DFPT.
The first-order change of density is computed from $u_{\textrm{d}m\textbf{k}}^{(1)}$, $f_{m\textbf{k}}^{(1)}$ and the zero-order quantities by

\begin{align}
\rho^{(1)}(\mathbf{r})
&= \, \, \, \,
n_{\textrm{s}}\!\int_{\textrm{BZ}}\!\sum_m 
f^{(1)}_{m\mathbf{k}}\, \rho^{(0)}_{m\mathbf{k}}(\mathbf{r}) \,[d\mathbf{k}] \nonumber \\
& \quad +
n_{\textrm{s}}\!\int_{\textrm{BZ}}\!\sum_m 
f^{(0)}_{m\mathbf{k}} \big(u^{(0)*}_{m\mathbf{k}}(\mathbf{r})\, u^{(1)}_{\textrm{d}m\mathbf{k}}(\mathbf{r})+\mathrm{c.c.}\big)\,[d\mathbf{k}].
\label{eq:rho1}
\end{align}
Eq.(\ref{eq:177}) is to be minimized by varying $u_{\textrm{d}m\textbf{k}}^{(1)}$
and
$f_{m\textbf{k}}^{(1)}$ under constraints
\begin{equation}
\langle u_{m\textbf{k}}^{(0)}| u_{\textrm{d}m\textbf{k}}^{(1)} \rangle=0
\end{equation}
and
 \begin{equation}
        n_\textrm{s} \int_{\textrm{BZ}} \sum_{m} 
         f_{m\textbf{k}}^{(1)}  
        [d\textbf{k}]=0.
        \label{eq:176}
    \end{equation}
Eq.(\ref{eq:176}) is equivalent to the condition of conservation of the number of particles for different temperatures: 
\begin{eqnarray}
\int_{\Omega_0} \rho^{(1)}(\textbf{r}) d\textbf{r}
=0.
\label{eq:int_rho1}
\end{eqnarray}

The minimization of Eq.(\ref{eq:177}) with respect to variations of the wavefunctions $| u_{m\textbf{k}}^{(1)} \rangle$ yields the  projected Sternheimer equation, that allows one to determine them,

\begin{eqnarray}
\hat{P}_{\perp m\textbf{k}}\Big( \hat{H}_{\textbf{k}\textbf{k}}^{(0)}
-\varepsilon_{m\textbf{k}}^{(0)}
\Big)
\hat{P}_{\perp m\textbf{k}}
| u_{m\textbf{k}}^{(1)} \rangle 
&=&
- 
\hat{P}_{\perp m\textbf{k}}\hat{H}^{(1)}
| u_{m\textbf{k}}^{(0)} \rangle,
\nonumber\\
\label{eq:Sterneimer_DeltaT_1}
\end{eqnarray}
where
$\hat{P}_{\perp m\textbf{k}}$ is the projector
on the subspace orthogonal to $|u_{m\textbf{k}}^{(0)} \rangle$. The first-order
Hamiltonian includes only the self-consistent
change of Hartree and exchange-correlation potential, $\hat{H}^{(1)}=\hat{v}_{\textrm{Hxc}}^{(1)}$,
that is local and $ \textbf{k}$-independent,
and originates from the modification
of the density:
\begin{equation}
v_{\textrm{Hxc}}^{(1)}(\textbf{r}) =
\Big(\hat{K}_{\textrm{Hxc}}^{(0)}
\rho^{(1)}
\Big)
(\textbf{r})
=
\int 
K_{\textrm{Hxc}}[\rho^{(0)}](\textbf{r},\textbf{r}')
\rho^{(1)}(\textbf{r}')
d\textbf{r}.
\label{eq:KHxc1}
\end{equation}
The Hartree and exchange-correlation kernel $K_{\textrm{Hxc}}[\rho](\textbf{r},\textbf{r}')$, functional of the density, is to be evaluated at $\rho^{(0)}$. 

The minimization of Eq.(\ref{eq:177}) with respect to variations of the occupation numbers 
$f_{m\textbf{k}}^{(1)}$
yields the following equation, that gives them directly:
\begin{equation}
f_{m\textbf{k}}^{(1)}=
\frac{\partial f}{\partial \varepsilon_{m\textbf{k}}}\Big|^{(0)}
\Big(
-\frac{\varepsilon^{(0)}_{m\textbf{k}}-\mu^{(0)}}{T^{(0)}}
+\varepsilon^{(1)}_{m\textbf{k}}-\mu^{(1)} \Big),
 \label{eq:147}
\end{equation}
where the following shorthand has been introduced,
\begin{equation}
\frac{\partial f}{\partial \varepsilon_{m\textbf{k}}}\Big|^{(0)}=
\frac{1}{k_\textrm{B}T^{(0)}}
\frac{\partial f_\textrm{FD}}{\partial x} 
\Bigg|_{
\frac
{\varepsilon_{m\textbf{k}}^{(0)}-\mu^{(0)}}
{k_\textrm{B}T^{(0)}}}
,
 \label{eq:dfFDde0}
\end{equation}
and $\frac{\partial f_\textrm{FD}}{\partial x}=-f_\textrm{FD}(x)(1-f_\textrm{FD}(x))$.
The first-order change of eigenenergy is obtained thanks to the Hellmann-Feynman theorem,
\begin{eqnarray}
\varepsilon_{n\textbf{k}}^{(1)}=
\langle u_{m\textbf{k}}^{(0)} |
\hat{H}^{(1)}
| u_{n\textbf{k}}^{(0)} \rangle,
\label{eq:Enk1}
\end{eqnarray}
and $\mu^{(1)}$ fixed by the constraint
Eq.(\ref{eq:176}) combined with Eq.(\ref{eq:147}):
\begin{widetext}
\begin{equation}
\mu^{(1)}=
\Big(
n_\textrm{s} \int_{\textrm{BZ}} \sum_{m}
\frac{\partial f}{\partial \varepsilon_{m\textbf{k}}}\Big|^{(0)}
\big(
-\frac{\varepsilon^{(0)}_{m\textbf{k}}-\mu^{(0)}}{T^{(0)}}
+\varepsilon^{(1)}_{m\textbf{k}} 
\big)
 [d\textbf{k}]
\Big)
\Big/
\Big(
n_\textrm{s} \int_{\textrm{BZ}} \sum_{m}
\frac{\partial f}{\partial \varepsilon_{m\textbf{k}}}\Big|^{(0)}
 [d\textbf{k}]
\Big).
\label{eq:mu1}
\end{equation}
\end{widetext}

Eq.(\ref{eq:147}) differs from the usual DFPT expression for the change of occupation number for metals, Eq.(63) of Ref.~\onlinecite{Gonze2024a}, by the presence 
of the term 
$-(\varepsilon^{(0)}_{m\textbf{k}}-\mu^{(0)})/T^{(0)}$ inside the parenthesis.
Indeed, a change of temperature induces a direct change of occupation numbers, while for the other types of perturbations, an 
occupation number is changed only indirectly, in response to the perturbation, due to the modification of 
its eigenvalue 
$\varepsilon_{n\textbf{k}}^{(1)}$ and the
modification of the chemical potential $\mu^{(1)}$. 
In the present case a non-self-consistent, ``bare'', occupation number change
is self-consistently  modified by an induced occupation number change.
The chemical potential change can be similarly decomposed, since the additional term 
$-(\varepsilon^{(0)}_{m\textbf{k}}-\mu^{(0)})/T^{(0)}$ is also present in the numerator of Eq.(\ref{eq:mu1}), needed to ensure the global charge neutrality of the response. The same fraction
is also the key signature of the perturbation in the second-order free energy expression Eq.(\ref{eq:177}).
These direct and induced modifications of $\mu^{(1)}$ are explicited as follows.
We detail the different components of  Eq.(\ref{eq:mu1}), and express them in terms of the DOS whenever possible. We define
\begin{align}
I\Big(\frac{\partial f}{\partial \varepsilon}\Big)&=
n_\textrm{s} \int_{\textrm{BZ}} \sum_{m}
\frac{\partial f}{\partial \varepsilon_{m\textbf{k}}}\Big|^{(0)}
 [d\textbf{k}]
 \nonumber
 \\
&=
\frac{1}{\Omega_0} \int_{-\infty}^{+\infty} 
\frac{\partial f}{\partial \varepsilon}\Big|^{(0)}
g_{\textrm{DOS}}^{(0)}(\varepsilon)
 d\varepsilon,
\label{eq:int_den}
\end{align}

\begin{align}
I\Big(\frac{\partial f}{\partial \varepsilon}\varepsilon\Big)&=
n_\textrm{s} \int_{\textrm{BZ}} \sum_{m}
\frac{\partial f}{\partial \varepsilon_{m\textbf{k}}}\Big|^{(0)}
\varepsilon_{m\textbf{k}}^{(0)}
 [d\textbf{k}]
 \nonumber
 \\
&=
\frac{1}{\Omega_0} \int_{-\infty}^{+\infty} 
\frac{\partial f}{\partial \varepsilon}\Big|^{(0)}
g_{\textrm{DOS}}^{(0)}(\varepsilon)
\varepsilon
 d\varepsilon,
\label{eq:int_den_eps}
\end{align}
and
\begin{align}
I\Big(\frac{\partial f}{\partial \varepsilon}\varepsilon^{(1)}\Big)&=
n_\textrm{s} \int_{\textrm{BZ}} \sum_{m}
\frac{\partial f}{\partial \varepsilon_{m\textbf{k}}}\Big|^{(0)}
\varepsilon_{m\textbf{k}}^{(1)}
 [d\textbf{k}].
\label{eq:int_den_eps1}
\end{align}
In the latter case, the integral cannot be straightforwardly 
changed from a Brillouin zone integration to an energy integration, as $\varepsilon_{m\textbf{k}}^{(1)}$ is not a simple function of the energy.

Eqs.(\ref{eq:int_den})-(\ref{eq:int_den_eps1}) allow one to write the bare
change of chemical potential, 
\begin{equation}
\mu_\textrm{bare}^{(1)}=
-
\frac
{I\Big(\frac{\partial f}
     {\partial \varepsilon}
     \varepsilon
     \Big)-\mu^{(0)}I\Big(\frac{\partial f}
     {\partial \varepsilon}
     \Big)}
{ T^{(0)}I\Big(\frac{\partial f}{\partial \varepsilon}\Big)},
\label{eq:mu1_NSC}
\end{equation}
and the induced one,
\begin{equation}
\mu_\textrm{ind}^{(1)}=
\frac
{I\Big(\frac{\partial f}
     {\partial \varepsilon}
     \varepsilon^{(1)}
     \Big)}
{ I\Big(\frac{\partial f}{\partial \varepsilon}\Big)},
\label{eq:mu1_scr}
\end{equation}
giving the total first-order change of chemical potential
\begin{equation}
\mu^{(1)}=
\mu_\textrm{bare}^{(1)}+
\mu_\textrm{ind}^{(1)}.
\label{eq:mu1_decomposition}
\end{equation}
The occupation number decomposition writes
\begin{equation}
f_{\textrm{bare},m\textbf{k}}^{(1)}=
\frac{\partial f}{\partial \varepsilon_{m\textbf{k}}}\Big|^{(0)}
\Big(
-\frac{\varepsilon^{(0)}_{m\textbf{k}}-\mu^{(0)}}{T^{(0)}}
-\mu_\textrm{bare}^{(1)} \Big),
 \label{eq:f1_NSC}
\end{equation}

\begin{equation}
f_{\textrm{ind},m\textbf{k}}^{(1)}=
\frac{\partial f}{\partial \varepsilon_{m\textbf{k}}}\Big|^{(0)}
\Big(
\varepsilon^{(1)}_{m\textbf{k}}
-\mu_\textrm{ind}^{(1)} \Big),
 \label{eq:f1_ind}
\end{equation}
giving the total first-order change of occupation number
\begin{equation}
f_{m\textbf{k}}^{(1)}=
f_{\textrm{bare},m\textbf{k}}^{(1)}
+
f_{\textrm{ind},m\textbf{k}}^{(1)}
.
 \label{eq:f1}
\end{equation}

One can check that the bare first-order change of occupation number fulfills the charge neutrality condition, and similarly for the induced one:
 \begin{equation}
        n_\textrm{s} \int_{\textrm{BZ}} \sum_{m} 
         f_{\textrm{bare},m\textbf{k}}^{(1)}  
        [d\textbf{k}]=0
        \label{eq:176NSC}
    \end{equation}
and
\begin{equation}
        n_\textrm{s} \int_{\textrm{BZ}} \sum_{m} 
         f_{\textrm{ind},m\textbf{k}}^{(1)}  
        [d\textbf{k}]=0.
        \label{eq:176ind}
    \end{equation}

The density change can be similarly decomposed in bare density change due to the modification of temperature, and the 
additional induced density change due to self-consistency. Explicitly,

\begin{equation}
\rho^{(1)}_\textrm{bare}(\mathbf{r})
=
n_{\textrm{s}}\!\int_{\textrm{BZ}}\!\sum_m 
f^{(1)}_{\textrm{bare},m\mathbf{k}}\, \rho^{(0)}_{m\mathbf{k}}(\mathbf{r}) \,[d\mathbf{k}],
\label{eq:rho1_NSC}
\end{equation}
and
\begin{align}
\rho^{(1)}_\textrm{ind}(\mathbf{r})
&= \, \, \, \,
n_{\textrm{s}}\!\int_{\textrm{BZ}}\!\sum_m 
f^{(1)}_{\textrm{ind},m\mathbf{k}}\, \rho^{(0)}_{m\mathbf{k}}(\mathbf{r})\,[d\mathbf{k}] \nonumber \\
& \quad +
n_{\textrm{s}}\!\int_{\textrm{BZ}}\!\sum_m 
f^{(0)}_{m\mathbf{k}} \big(u^{(0)*}_{m\mathbf{k}}(\mathbf{r})\, u^{(1)}_{\textrm{d}m\mathbf{k}}(\mathbf{r})+\mathrm{c.c.}\big)\,[d\mathbf{k}].
\label{eq:rho1ind}
\end{align}

Connecting the induced density change to the change of potential proceeds now similarly to the analysis of the SCF cycle for periodic metals performed in
Ref.\onlinecite{Herbst2020a}.
The independent-particle susceptibility $\chi_0$
connecting both of them is introduced,
\begin{equation}
\delta \rho(\textbf{r}) =
\Big(\hat{\chi}_0\delta v\Big)(\textbf{r})
=
\int 
\chi_0(\textbf{r},\textbf{r}')
\delta v(\textbf{r}')
d\textbf{r}'.
\label{eq:drho_dv}
\end{equation}
with an Adler-Wiser contribution and a Fermi contribution to $\hat{\chi}_0$.
The Adler-Wiser change of density is given by
\begin{equation}
\Big(\delta \rho(\textbf{r})\Big)_{\textrm{AW}} =
\Big(\hat{\chi}_{\textrm{AW}}\delta v\Big)(\textbf{r})
=
\int 
\chi_{\textrm{AW}}(\textbf{r},\textbf{r}')
\delta v(\textbf{r}')
d\textbf{r}'.
\label{eq:drho_dv_AW}
\end{equation}
The kernel of this operator can be obtained from a sum over states and a double integral over the Brillouin Zone, as follows,
\begin{align}
\chi_{\textrm{AW}}(\textbf{r},\textbf{r}')
= 
n_{\textrm{s}}
\Omega_0^2
\int_{\textrm{BZ}}
&
\int_{\textrm{BZ}}
\sum_{nm}
\frac
{\occf_{m\textbf{k}+\textbf{q}}-\occf_{n\textbf{k}} }{
\varepsilon_{m\textbf{k}+\textbf{q}}-\varepsilon_{n\textbf{k}} 
}
\nonumber\\
\exp(-i\textbf{q}(\textbf{r}'-\textbf{r}))
&
M^{mn*}_{\textbf{k}+\textbf{q},\textbf{k}}(\textbf{r})
M^{mn}_{\textbf{k}+\textbf{q},\textbf{k}}(\textbf{r}')
[d\textbf{k}]
[d\textbf{q}],
\label{eq:chi0}
\end{align}
with
%
\begin{equation}
M^{mn}_{\textbf{k}+\textbf{q},\textbf{k}}(\textbf{r}) =
u_{m\textbf{k}+\textbf{q}}^* (\textbf{r})
u_{n\textbf{k}}(\textbf{r}).
\label{eq:Mmn}
\end{equation}
In this expression, the ratio between differences of occupation numbers and differences of energies must be treated carefully, with limiting behavior
\begin{align}
\lim_{(\varepsilon_{m\textbf{k}+\textbf{q}}-\varepsilon_{n\textbf{k}})\rightarrow0}
\frac
{\occf_{m\textbf{k}+\textbf{q}}-\occf_{n\textbf{k}} }{
\varepsilon_{m\textbf{k}+\textbf{q}}-\varepsilon_{n\textbf{k}} 
}
=\frac{\partial{f}}{\partial \varepsilon}
\Big|_{\varepsilon_{n\textbf{k}}}.
\label{eq:limit_f_e}
\end{align}
Also, the treatment of the thermodynamic limit must be done carefully, and we refer the reader to Sec. IV A of Ref.\onlinecite{Herbst2020a} on this matter.

The analysis of the self-consistent DFT behavior is done thanks to the electronic dielectric operator 
\begin{equation}
\hat{\epsilon}_{\textrm{e}}=\hat{1}-\hat{K}_{\textrm{Hxc}}\hat{\chi}_0,
\label{eq:elec_diel}
\end{equation}
where, in the present context, $\hat{\chi}_0$ is evaluated at finite temperature. 
The inverse of $\hat{\epsilon}_{\textrm{e}}$ screens the bare potential change, to deliver the self-consistent one. Similarly, its inverse transpose screens the density change. 
Indeed,
one derives from Eqs.(\ref{eq:KHxc1}), (\ref{eq:rho1_NSC}),
(\ref{eq:rho1ind}),
(\ref{eq:drho_dv})
and (\ref{eq:elec_diel}), 
the relation between the self-consistent change of density and the bare one,
\begin{eqnarray}
\rho^{(1)}(\textbf{r}) &=& 
\int
\epsilon_{\textrm{e}}^{-1t}(\textbf{r},\textbf{r}')
\rho^{(1)}_{\textrm{bare}}(\textbf{r}') 
d\textbf{r}'. 
\label{eq:DTn2_1}
\end{eqnarray}

As a final result in this section, one can derive generic formulas for the second-order mixed derivatives of the free energy per unit cell with respect to two perturbations: the temperature change and another, generic, one, whose small expansion parameter is denoted by $\lambda$.
Building upon the result Eq.(\ref{eq:F1el}), one gets

\begin{align}
\frac{\partial^2 F_\textrm{el}}
{\partial T \partial \lambda}
\Big|_{T^{(0)},\lambda=0}  
&=
-\frac{\partial S\big(\{f_{n\textbf{k}}(\lambda)\delta_{mn}\}\big)
}{\partial \lambda}\Big|_{T^{(0)},\lambda=0}
\nonumber\\
=
-&
n_{\textrm{s}}\Omega_0
\int_\textrm{BZ}
\sum_n
k_\textrm{B}
\frac{\partial s_\textrm{FD}
}{\partial \lambda}
\Big|_{T^{(0)},\lambda=0}
[d\textbf{k}]
\nonumber\\
=
-&
n_{\textrm{s}}\Omega_0
\int_\textrm{BZ}
\sum_n
\frac
  {\varepsilon_{n\textbf{k}}^{(0)}-\mu^{(0)}}
  {T^{(0)}}
\frac
  {\partial f_{n\textbf{k}}}
  {\partial \lambda}
\Big|_{T^{(0)},\lambda=0}
[d\textbf{k}]
.\label{eq:F2mixed}
\end{align}
For the final step in this derivation, see Eq.(10) of Ref.\onlinecite{Gonze2024a}.
In this expression, neither the derivative of the wavefunction with respect to the temperature,
nor the one of the occupation numbers appear.
The only first-order ingredients are the first-order
derivatives of the occupation numbers with respect
to the generic perturbation. 

The second-order mixed derivatives of the free energy per unit cell with respect to two perturbations, one being the temperature, can
also be computed thanks to first-order derivatives
with respect to the temperature, without using the
first-order derivatives with respect to the other
perturbation. Such expressions depends on the specific form of the generic perturbation. 
For example, supposing one deals
with a change of external potential, then
\begin{widetext}
\begin{align}
\frac{\partial^2 F_\textrm{el}}
{\partial T \partial \lambda}
\Big|_{T^{(0)},\lambda=0}=
n_{\textrm{s}} \Omega_0
\int_{\textrm{BZ}}
\sum_{n}
\Bigg(
\frac{\partial f_{n\textbf{k}}}{\partial T} 
\langle u_{\textrm{d}n\textbf{k}}^{(0)} | 
\frac{\partial \hat{v}_{\textrm{ext},\textbf{k}\textbf{k}}}{\partial \lambda} 
| u_{\textrm{d}n\textbf{k}}^{(0)} \rangle
+
f_{n\textbf{k}}
\Big(
\langle 
\frac{\partial u_{\textrm{d}n\textbf{k}}}{\partial T} | \frac{\partial \hat{v}_{\textrm{ext},\textbf{k}\textbf{k}}}{\partial \lambda} | u_{\textrm{d}n\textbf{k}}^{(0)} \rangle
+\textrm{(c.c.)}\Big)
\Bigg)
[d\textbf{k}].
\label{eq:F2mixed2}
\end{align}
\end{widetext}

In this expression, neither the derivative of the wavefunction with respect to the generic perturbation,
nor the one of the occupation numbers appear.
The only first-order ingredients are the first-order
derivatives with respect to the temperature and
the derivative of the external potential, which is actually given in the definition of the perturbation. 

When the reference temperature vanishes, several equations above are well-behaved. For example, the derivative $\partial f/\partial \varepsilon$ in Eq.(\ref{eq:147}) in the limit $T^{(0)}\to 0$ is to be replaced by 
\begin{equation}
\frac{\partial f}{\partial \varepsilon}\Big|^{(0)}
\;\xrightarrow[T^{(0)}\to 0]{}\;
-\delta\big(\varepsilon_{m\mathbf{k}}^{(0)}-\mu^{(0)}\big),
\label{eq:dfde_T0}
\end{equation}
that behaves properly when introduced in Brillouin Zone integrals Eqs.(\ref{eq:int_den}), 
(\ref{eq:int_den_eps}), (\ref{eq:int_den_eps1}), 
(\ref{eq:int_den_eps1}),
(\ref{eq:176NSC}), and
(\ref{eq:176ind}).

However, the presence of $1/T^{(0)}$ in the first term of the rightmost factor in Eq.(\ref{eq:147})
(also present in Eqs.(\ref{eq:177}),(\ref{eq:mu1}) and (\ref{eq:F2mixed})) 
triggers a more serious challenge, and will be addressed thanks to the
Sommerfeld expansion, as we shall see in Secs.~\ref{sec:Sommerfeld}
and~\ref{sec:lowT-DFT}.

\section{Validation of the DFPT treatment of a temperature change}
\label{sec:Validation}
The perturbation of the solution of the DFT equations with respect to
temperature falls naturally into the framework of algorithmic
differentiation as implemented in the DFTK code
\cite{herbst2021dftk,schmitz2025algorithmic}. In DFTK, calculations
are split into (1) the preparation phase of building the model,
putting all relevant input parameters (in this particular instance,
the temperature) into the appropriate data structures, (2) solving the Kohn-Sham equations for the
electronic degrees of freedom, and (3) postprocessing (in this case,
computing the energy and entropy). Differentiation of the first and
third phases is performed in an algorithmic way by the ForwardDiff
Julia type \cite{revels2016forward}, with only the ``core'' part of
teaching the algorithm how to differentiate through the Kohn-Sham
equations using the Dyson and Sternheimer equations of DFPT coded by
hand. In particular, since this core part is agnostic to the input and
output, starting from the existing implementation of DFPT in
DFTK, no specific code needed to be added to compute the derivative
with respect to temperature (modulo bug fixing and taking care of
numerically unstable calculations like computing the derivative of the
entropy function).

\begin{figure}[t]
  \centering
  \includegraphics[width=\columnwidth]{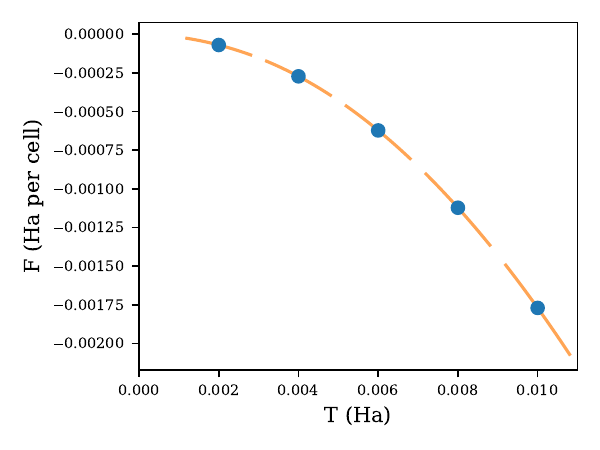}
  \caption{The free energy per unit cell of aluminium  as a function of
    Fermi-Dirac temperature, taking as reference the (extrapolated)
    $T=0$ result. The branches of parabolas
    are computed using free-energy values, and first- and second- derivative information obtained from the
  method described in the text. Quantities are given in atomic unit, with k$_B$ taken equal to 1. The conversion factor to Kelvin is 0.010Ha=3157.77K, corresponding to the highest temperature shown.}
  \label{fig:Al_1}
\end{figure}

\begin{figure}[t]
  \centering
  \includegraphics[width=\columnwidth]{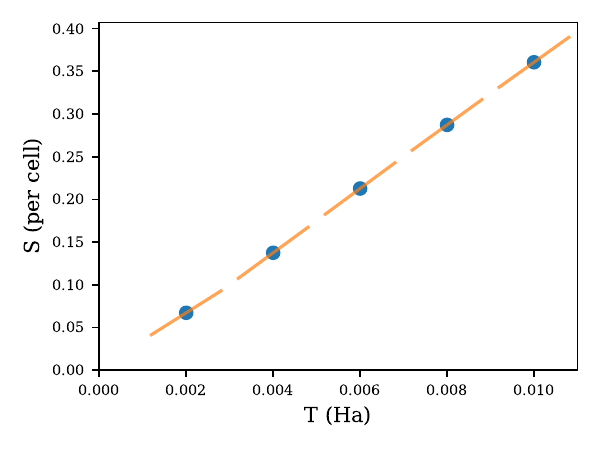}
  \caption{The entropy per unit cell of aluminium as a function of
    Fermi-Dirac temperature. The segments are computed
    using the first derivative information obtained from the method
    described in the text. Quantities are given in atomic unit, with k$_B$ taken equal to 1. The conversion factor to Kelvin is 0.010Ha=3157.77K, corresponding to the highest temperature shown.}
  \label{fig:Al_2}
\end{figure}

We tested this on FCC Aluminum, using the same parameters as in
Ref.\onlinecite{Verstraete2004}: a small plane wave energy cut-off of 10 Ha, and a
Monkhorst-Pack grid of $26^{3}$ points. We obtain the free energy and
its first derivative, $\frac{dF}{dT}=-S$, from a standard DFT calculation. Then we use
the DFPT automatic differentiation to obtain $\frac{d^2F}{dT^2}=-\frac{dS}{dT}$. The results are presented in
Figs \ref{fig:Al_1} and \ref{fig:Al_2}.

In Fig.\ref{fig:Al_1}, for five different temperatures, the free energy as well as a second-order parabola from the first- and second-order derivatives of the free energy with respect to temperature have been represented, with excellent match between
the different parabolas.
Fig.\ref{fig:Al_2} similarly presents the entropy as well as its first-order derivative.
For all temperatures, the values of the entropy corresponds very well to the slopes of \ref{fig:Al_1}, as expexted.
At the highest temperatures, the slopes of the entropy are well aligned with a global linear behavior.
However, for $T=0.002$Ha, the slope of the entropy does not exactly yield a zero entropy for zero temperature, while it should, according to Nernst principle (third law of thermodynamics).
This originates from the wavevector sampling grid: we have checked that the discrepancy is significantly decreased by increasing the sampling from 12$^3$ to 40$^3$. The 26$^3$ grid was kept to illustrate this difficulty, but is not sufficiently dense for the smallest of our temperatures. This
phenomenon can be related to an inaccurate evaluation of the density of state at the Fermi energy.
Such numerical integration problem must be taken into account to obtain accurate results at low temperatures, in addition to the already mentioned $1/T^{(0)}$ divergences. 
Improving the grid sampling allows one to fix the issue. Alternatively, it is also possible
to combine the Fermi-Dirac 
smearing, that is directly determined by the temperature, with a resmearing technique, as mentioned
in Ref.\onlinecite{Verstraete2001, Verstraete2004, Gonze2024a}.

We now address the low-temperature limit using the Sommerfeld expansion
in order to avoid the $1/T^{(0)}$ divergences.
We do not further focus on the issue of wavevector grid sampling.


\section{The Sommerfeld expansion}
\label{sec:Sommerfeld}

In this section, results obtained from the Sommerfeld expansion for the non-interacting free electron gas of (homogeneous) density $\rho$, are first recalled: following Sommerfeld, the lowest-order temperature dependence of the chemical potential and internal energy are obtained\cite{Sommerfeld1928}.
The mathematical characterization of the convergence of the Sommerfeld expansion, that is not a Taylor expansion, is outlined.
Then, this section continues with 
results for the usage of the Sommerfeld expansion beyond the free electron gas, and finishes with
results for the case of Van Hove singularities.

\subsection{Low-temperature free electron gas}

In the free electron gas, each electronic state has an eigenenergy 
\begin{equation}
\label{eq:ek_FEG}
\varepsilon_{\textbf{k}}=\frac{\textbf{k}^2}{2m^*},
\end{equation}
where $m^*$ is the electronic effective mass.
The corresponding eigenfunction is a simple planewave.
The occupation of each electronic state is governed by the temperature-dependent Fermi-Dirac occupation , but the dispersion relation Eq.(\ref{eq:ek_FEG}) is independent of temperature.
 
The electron gas density  $\rho$ is inferred from a simplified version of Eq.~(\ref{eq:n_metal_diagonal}),
\begin{eqnarray}
\rho(T) = 
n_{\textrm{s}}
\int
\occf_{\textbf{k}}(T,\mu)
[d\textbf{k}].
\label{eq:n_FEG}
\end{eqnarray}
It is homogeneous, without dependence on $\mathbf{r}$.
Working at constant number of particles, instead of fixed chemical potential 
$\mu$, the electron gas density $\rho$
is kept fixed as a function of the temperature.
Then, from Eq.~(\ref{eq:n_FEG}), $\mu$ becomes a function of $T$ for a fixed $\rho$.
The internal energy density $E(T)$ (energy per unit volume) is given by 
\begin{eqnarray}
E(T) = 
n_{\textrm{s}}
\int
\varepsilon_{\textbf{k}}
\occf_{\textbf{k}}(T,\mu(T))
[d\textbf{k}],
\label{eq:internalE_FEG}
\end{eqnarray}
where the dependence of $\mu$ on $T$
has been made explicit.

The free electron gas DOS is independent of the temperature, unlike in 
Eq.(\ref{eq:gDOS_e_inhom_1})):
\begin{eqnarray}
g_{\textrm{DOS}}(\varepsilon)= 
n_{\textrm{s}}
\int
\delta(\varepsilon_{\textbf{k}}-\varepsilon)
[d\textbf{k}].
\label{eq:gDOS_e_homogeneous}
\end{eqnarray}
In three dimensions,  one obtains~\cite{Sommerfeld1928}:
$g_{\textrm{DOS}}^{3\textrm{D}}(\varepsilon) = 0$ when $\varepsilon < 0 $, while
when $\varepsilon \geq 0$,
\begin{equation}
    g_{\textrm{DOS}}^{3\textrm{D}}(\varepsilon)
    =C^{\textrm{3D}}
    \varepsilon^{\frac{1}{2}}, 
    \textrm{ where }C^{\textrm{3D}}=\Big( 2(m^{*})^{3}\Big)^{\frac{1}{2}},
\end{equation}
with a characteristic square-root dependency on the energy. 
It is thus continous and derivable at all orders, except at 
$\varepsilon=0$, where a Van Hove singularity happens, with the typical 3D
square-root dependency\cite{VanHove1953} of a parabolic band extremum.

The DOS allows one to rewrite the 
$\textbf{k}$-integral of a 
generic $\varepsilon_{\textbf{k}}$-dependent function, $\alpha(\varepsilon_{\textbf{k}})$,
as
\begin{align}
I&=n_{\textrm{s}}
\int
\alpha(\varepsilon_{\textbf{k}})
[d\textbf{k}]=
\int_{-\infty}^{\infty}
\alpha(\varepsilon)
n_{\textrm{s}}
\int
\delta(\varepsilon_{\textbf{k}}-\varepsilon)
[d\textbf{k}]
d\varepsilon
\nonumber
\\
&=
\int_{-\infty}^{\infty}
\alpha(\varepsilon)
g_{\textrm{DOS}}(\varepsilon)
d\varepsilon.
\label{eq:k-integral}
\end{align}

Accordingly, the density and free energy are written as energy integrals,
\begin{eqnarray}
\rho
&=&
\int_{-\infty}^{\infty}
f_{\textrm{FD}}\bigg(\frac{\varepsilon-\mu(T)}{k_\textrm{B}T} \bigg)
C^{\textrm{3D}}\varepsilon^{\frac{1}{2}}
d\varepsilon,
\label{eq:rhoT_heg}
\end{eqnarray}
and
\begin{eqnarray}
E(T)
&=&
\int_{-\infty}^{\infty}
f_{\textrm{FD}}\bigg(\frac{\varepsilon-\mu(T)}{k_\textrm{B}T} \bigg)
C^{\textrm{3D}}\varepsilon^{\frac{3}{2}}
d\varepsilon.
\label{eq:ET_heg}
\end{eqnarray}

At this stage, the low-temperature expansion of the
integral over the energy is performed, using
Sommerfeld's result, see Eq.(26a) of the original work~\cite{Sommerfeld1928} and the section SuppMat3:
\begin{eqnarray}
I(T)
&=&
\int_{-\infty}^{\infty}
f_{\textrm{FD}}\bigg(\frac{\varepsilon-\mu(T)}{k_\textrm{B}T} \bigg)
h(\varepsilon)
d\varepsilon
\nonumber\\
&=&
\int_{-\infty}^{\mu} h(\varepsilon) d\varepsilon + \frac{\pi^{2}}{6}h^{'}(\mu)(k_\textrm{B}T)^{2} + {\cal{O}}(k_\textrm{B}T)^{4},
\nonumber
\\
\label{eq:sommerfeld-integral-up-to-2}
\end{eqnarray}
where $h^{'}(\mu)$ denotes the derivative of
the $h$ function with respect to $\varepsilon$ evaluated at $\mu$.
This result is valid provided that the
$h$ function is derivable twice. 

This delivers
\begin{align}
\rho
&\approx
\int_{-\infty}^{\mu(T)}
C^{\textrm{3D}}\varepsilon^{\frac{1}{2}}
d\varepsilon
+ \frac{\pi^{2}}{12}
C^{\textrm{3D}}
\Big(\mu(T)\Big)^{-\frac{1}{2}}
(k_\textrm{B}T)^{2}
\nonumber\\
&\approx
\frac{2}{3}
C^{\textrm{3D}}\big(\mu(T)\big)^{\frac{3}{2}}
+ \frac{\pi^{2}}{12}
C^{\textrm{3D}}
\Big(\mu(T)\Big)^{-\frac{1}{2}}
(k_\textrm{B}T)^{2},
\label{eq:rho_Sommerfeld}
\end{align}
yielding, up to quadratic order,
\begin{eqnarray}
\mu(T)\approx \varepsilon_\textrm{F}\bigg(
1-\frac{\pi^{2}}{12}\Big(\frac{k_\textrm{B}T}{\varepsilon_\textrm{F}}\Big)^2
\bigg),
\label{eq:muT_Sommerfeld}
\end{eqnarray}
where the Fermi energy $\varepsilon_\textrm{F}$ is defined as the chemical potential at zero temperature,
\begin{equation}
\varepsilon_\textrm{F}=\mu(T=0)
=\Big(\frac{3\rho}{2C^{\textrm{3D}}}
\Big)^{2/3}.
\end{equation}
The internal energy is obtained similarly. At quadratic order,
\begin{align}
E(T)
&\approx
\int_{-\infty}^{\mu(T)}
C^{\textrm{3D}}\varepsilon^{\frac{3}{2}}
d\varepsilon
+ \frac{\pi^{2}}{4}
C^{\textrm{3D}}
\Big(\mu(T)\Big)^{\frac{1}{2}}
(k_\textrm{B}T)^{2}
\nonumber
\\
&\approx
\frac{2}{5}C^{\textrm{3D}}
\Big(\mu(T)\Big)^{\frac{5}{2}}
+ \frac{\pi^{2}}{4}
C^{\textrm{3D}}
\Big(\mu(T)\Big)^{\frac{1}{2}}
(k_\textrm{B}T)^{2}
\nonumber
\\
&\approx
\varepsilon_\textrm{F} \rho
\Bigg(
\frac{3}{5}
+ \frac{\pi^{2}}{4}
\Big(\frac{k_\textrm{B}T}{\varepsilon_\textrm{F}}\Big)^2
\Bigg).
\label{eq:ET_Sommerfeld}
\end{align}

\subsection{Convergence of the Sommerfeld expansion}

Eqs.(\ref{eq:muT_Sommerfeld}) and (\ref{eq:ET_Sommerfeld}) have the characteristic form of a constant plus a $T^2$ contribution. The section SuppMat3 gives some background about the Sommerfeld expansion,
that is indeed an expansion in even powers of $T$ only.
For smooth (infinitely differentiable) functions, the coefficients of terms with
powers higher than 2 can be obtained exactly, in terms of Bernoulli numbers of order $n$, $B(n)$.
However, the convergence properties of the Sommerfeld expansion differ from  Taylor series of usual mathematical functions, as we describe now.

The Sommerfeld expansion is usually an \textit{asymptotic} series, and does
not converge for any finite value of $T$. In fact, there is a simple
characterization of the class of functions that make the Sommerfeld
expansion converge:
\begin{theorem}
  Let $f:\mathbb{R}\to \mathbb{C}$ be $C^{\infty}$ and $\mu \in \mathbb R$, and consider the formal
  Sommerfeld series
  \begin{align}
    Somm(T) = \sum_{n=1}^\infty \frac{M_{\textrm{FD}}(2n)}{(2n)!} h^{(2n-1)}(\mu) (k_{\rm B} T)^{2n},
    \label{eq:Somm}
  \end{align}
  where
  \begin{equation}
    M_{\textrm{FD}}(2n)=(-1)^{(n-1)}2 (\pi)^{2n} (2^{2n-1}-1)B_{2n},
  \end{equation}
and
$h^{(2n-1)}$ is the $2n-1$ derivative of the function $h$ evaluated at $\mu$.

Then $Somm(T)$ has a {nonzero radius of convergence} as a power series in
$T$ {if and only if} the odd part of $h$ about $\mu$, $h_{o}(\mu+x) = \tfrac 1 2 (h(\mu+x)-h(\mu-x))$ extends to an entire function of exponential type; that is, there exist constants $C,A>0$ such that
\[
|h_{o}(z)| \le C e^{A|z|}, \qquad \forall z\in\mathbb{C}.
\]
\end{theorem}
Taking the odd part is necessary here because the Sommerfeld expansion
only sees the odd derivatives of $h$ at $0$. The proof uses standard
arguments but we were unable to find an explicit reference, so
we reproduce it in the section SuppMat4.

The class of functions of exponential type is rather small, and in
particular rules out most of the functions $h$ of practical interest,
such as density of states, that have non-analyticities on the real
axis. Therefore, the Sommerfeld expansion only contains information on
the behavior of
$\int_{-\infty}^{\infty} h(\varepsilon) f_{\rm FD}((\mu-\varepsilon)/(k_{\rm B}T))$
for infinitesimally small $T$, not at any finite $T$ - this can also
be seen intuitively by noticing that any modification of $h$ away from
$\mu$ (for instance, adding an extra band to the model) results in a
contribution exponentially small in $T$, which cannot be picked up by
a power series.

In practice, however, the Sommerfeld expansion is doing rather well for typical well-behaved metals: it explains the experimentally found linear behavior of the specific heat of many metals. Also the first-principles quadratic behavior
of the free energy of many metals is indeed common. 
Finally, the first-principles quadratic behavior of other properties of metals, like the square of phonon frequencies, has also been 
noticed~\cite{Tantardini2025a}.

\subsection{Beyond the free electron gas}

The more realistic case of
lattice periodic metals, with several bands and general dispersion relations, is considered now, including the interaction between electrons.
Due to self-consistency, the dispersion relation is temperature-dependent, the DOS similarly, while the chemical potential depends on the temperature in order to keep constant the integrated charge.
Also, in this subsection, the mathematical considerations of the previous two subsections 
are enriched to deal with entropy.

In order for the Sommerfeld expansion 
Eq.(\ref{eq:sommerfeld-integral-up-to-2}) to be used, 
the electronic density, the internal energy, the entropy and the 
free energy, initially expressed using
Brillouin Zone integrals, are formulated in terms of energy integrals.
This reformulation has already been done in Sec.~\ref{Sec:VDFTM} 
for the electronic density,
compare Eqs.(\ref{eq:n_metal_diagonal}) and (\ref{eq:n_metal_int_e_1}).

The number of electrons per unit cell, Eq.(\ref{eq:N_e}) is reformulated as follows, using Eqs.(\ref{eq:n_metal_int_e_1}) and (\ref{eq:gDOS_e_inhom_1}):
\begin{eqnarray}
N_e = 
\int_{-\infty}^{+\infty}
f_{\textrm{FD}}\bigg(\frac{\varepsilon-\mu(T)}{k_\textrm{B}T} \bigg)
g_{\textrm{DOS}}(T,\varepsilon)
d\varepsilon . 
\label{eq:Ne_energy}
\end{eqnarray}

The internal energy per unit cell, in the non-interacting case, is 
\begin{eqnarray}
E_{\textrm{NI}}(T) = 
n_{\textrm{s}}\Omega_0
\int_{\textrm{BZ}}
\varepsilon_{n\textbf{k}}
f_{n\textbf{k}}(T,\mu(T))
[d\textbf{k}],
\label{eq:eq:EniT_k}
\end{eqnarray}
and becomes
\begin{eqnarray}
E_{\textrm{NI}}(T) =
\int_{-\infty}^{+\infty}
\, \varepsilon \,
f_{\textrm{FD}}\bigg(\frac{\varepsilon-\mu(T)}{k_\textrm{B}T} \bigg)
g_{\textrm{DOS}}(T,\varepsilon)
d\varepsilon . 
\label{eq:EniT_energy}
\end{eqnarray}

In order to obtain the internal
energy per unit cell in the DFT case, that includes the Hartree and exchange-correlation contribution, corrections terms that depend only on the density are added:
\begin{eqnarray}
E(T)
=E_{\textrm{NI}}(T)
+ 
E_{\textrm{Hxc}}[\rho] 
-
\int_{\Omega_0}\rho(\textbf{r}) v_{\textrm{Hxc}}[\rho]
d\textbf{r}.
\label{eq:E=ENI+Hxc}
\end{eqnarray}

The entropy per unit cell, 
Eq.(\ref{eq:sFD}), formulated in terms of the occupation numbers, is given by
\begin{equation}
S(T) = 
n_{\textrm{s}}\Omega_0
\int_{\textrm{BZ}}
k_\textrm{B}s_\textrm{FD}\big(f_{n\textbf{k}}(T,\mu(T))\big)
[d\textbf{k}].
\label{eq:entropy}
\end{equation} 
The internal energy per unit cell and the entropy per unit cell combine to deliver the free energy per unit cell,
\begin{equation}
F(T)=E(T)-TS(T).
\label{eq:FT}
\end{equation}

Unlike the electronic density, 
the number of electrons per unit cell and the internal energy per unit cell, it is not obvious
how the entropy per unit cell, Eq.(\ref{eq:entropy}),
can be rewritten as an integral of the FD occupation function multiplying a function of the energy.
At variance, it can be expressed in terms of the derivative of minus the FD occupation function $f_\textrm{FD}(x)$, Eq.(\ref{eq:FD}), 
denoted $\Tilde{\delta}_\textrm{FD}(x)$,
\begin{align}
    \Tilde{\delta}_{\textrm{FD}}(x) =
    - \frac{df_\textrm{FD}}{dx}\Big|_x
    =
    \frac{1}{\big( \exp(x)+1\big)\big(\exp(-x)+1\big)},
    \label{eq:187}
\end{align} 
as we will see.
On the basis of such expression,
a series expansion can be derived,
similar to the Sommerfeld one.
The function $\Tilde{\delta}_\textrm{FD}(x)$ is normalized to one (see the section SuppMat3, $M_{\textrm{FD}}(0)=1$), and is peaked around $x=0$, thus it is a smeared Dirac delta function.
Note,
\begin{equation}
f_{\textrm{FD}}(x) = \int_{x}^{\infty} \Tilde{\delta}_{\textrm{FD}}(y) dy. \label{eq:185}
\end{equation}

For the purpose of computing $S(T)$, we will follow the 
 scheme introduced by Methfessel and coworkers in
Ref.~\onlinecite{Methfessel1989}. It delivers the entropy associated with the FD occupation function
in term of an integral of it. By the way, it also allows one to obtain the entropy associated to other occupation functions (this has been used in several recent studies to examine the effect of alternative numerical smearing schemes 
\cite{DosSantos2023, Gonze2024a}).

Methfessel \textit{et al.} define the auxiliary
entropy function $\Tilde{s}_\textrm{FD}(x)$, a symmetric
adimensional function of an adimensional argument $x$,
\begin{equation}
\Tilde{s}_\textrm{FD}(x) = \int_{x}^{ \infty} y \Tilde{\delta}_\textrm{FD}(y) dy=
-\int^{x}_{-\infty} y \Tilde{\delta}_\textrm{FD}(y) dy.
\label{eq:186} 
\end{equation}
From this expression, $S(T)$ is obtained as an integral, where the electronic energies appear instead of the occupation numbers,
\begin{equation}
S(T) = 
n_{\textrm{s}}\Omega_0
\int_{\textrm{BZ}}
k_\textrm{B}
\Tilde{s}_\textrm{FD}
\Big(
\frac{\varepsilon_{n\textbf{k}}-\mu(T)}
{k_\textrm{B}T}
\Big)
[d\textbf{k}].
\label{eq:entropy_2}
\end{equation}
It is then transformed to an energy integral,
\begin{equation}
S(T) = 
k_\textrm{B}
\int_{-\infty}^{\infty}
\Tilde{s}_\textrm{FD}
\Big(
\frac{\varepsilon-\mu(T)}
{k_\textrm{B}T}
\Big)
g_{\textrm{DOS}}(\varepsilon)
d\varepsilon,
\label{eq:entropy_3}
\end{equation}
and formulated as an integral that includes 
$\Tilde{\delta}_{\textrm{FD}}(x)$, thanks to 
Eq.(\ref{eq:186}),
\begin{eqnarray}
S(T) &=& 
k_\textrm{B}
\int_{-\infty}^{\infty}
\Bigg( \int^{\infty}_{\frac{\varepsilon-\mu(T)}
{k_\textrm{B}T}} x \Tilde{\delta}_\textrm{FD}(x) dx
\Bigg)
g_{\textrm{DOS}}(\varepsilon)
d\varepsilon.
\nonumber\\
\label{eq:entropy_deltaFD}
\end{eqnarray}
As shown in the section SuppMat3, after expanding the 
$g_{\textrm{DOS}}$ around $\varepsilon_\textrm{F}$,
the following low-order term in the series of $S(T)$ is obtained, in the case where the DOS is smooth:
\begin{align}
    S(T)   
&=  \frac{\pi^{2}}{3} k_\textrm{B}^{2} T g_{\textrm{DOS}} (\varepsilon_\textrm{F}) 
+ {\cal{O}}(k_\textrm{B}T)^{3}.
\label{eq:s63} 
\end{align}
For higher-order terms, see the
section SuppMat3.
Eq.(\ref{eq:s63}) is in line with Nernst principle: at $0K$, the entropy vanishes.

The same expansion might be obtained from the standard grand-canonical result for the ideal Fermi gas,\cite{Mahan2000,FetterWalecka2018}
\begin{align}
\Omega=-k_\mathrm{B}T \int d\varepsilon\, g_{\textrm{DOS}}(\varepsilon)\,
\ln\!\bigl(1+e^{(\mu-\varepsilon)/(k_\mathrm{B}T)}\bigr),
\end{align} 
together with Eq.(\ref{eq:Ne_energy}) for $N_e$ and the thermodynamic identity
relating Helmholtz free energy and grand potential,
\begin{equation}
F=\Omega+\mu N_e ,
\end{equation}
see, e.g., Ref.~\cite{Huang1987} and Ref.~\cite[Ch.~2--3]{ashcroft1976}.
This yields the working expression for $F(T)$,
\begin{align}
F(T) &= \int_{-\infty}^\infty \Big[\mu(T)\, f_\mathrm{FD}\!\left(\tfrac{\varepsilon-\mu(T)}{k_\textrm{B}T}\right) \nonumber \\
&- k_\textrm{B}T\ln\!\Big(1+e^{(\mu(T)-\varepsilon)/k_\textrm{B}T}\Big)\Big] g_{\textrm{DOS}}(\varepsilon)\, d\varepsilon,
\label{eq:numerics-FS}
\end{align}
then using 
Eq.(\ref{eq:F1=-S}).
The lowest-order free energy expansion is
\begin{align}
F(T)-F(0)=  
&-  \frac{\pi^{2}}{6} k_\textrm{B}^{2} T^2 g_{\textrm{DOS}} (\varepsilon_\textrm{F}) 
+ {\cal{O}}(k_\textrm{B}T)^{4}.
\label{eq:85} 
\end{align}

\subsection{Van Hove singularities}

From Eqs.(\ref{eq:Ne_energy}),
(\ref{eq:EniT_energy})
and (\ref{eq:entropy_deltaFD}), one sees that the DOS enters the Fermi-Dirac integrals that are
to be evaluated.
Be it for the homogeneous electron gas or the inhomogeneous systems, 
the DOS
is not smooth everywhere, so such factor does not meet the conditions for the application of Theorem 1. 
Still, at almost all energies\cite{gerard1990resonance}, the needed derivatives of the function exist, and the series Eq.(\ref{eq:Somm}) 
can be written, even if its
radius of convergence vanishes.
There are exception, the Van Hove singularities.
The DOS exhibits indeed Van Hove singularities for eigenenergies at which the gradient of the electronic eigenvalues as a function of the wavevector vanishes\cite{VanHove1953}. 
Singularities in the DOS might also appear even with non-vanishing gradient, if band crossings such as Dirac or Weyl  are present.
However we will not examine such cases and treat only the
usual Van Hove singularities linked
to a non-degenerate energy band with a vanishing gradient and a Hessian without zero eigenvalue at the critical energy.

At the corresponding energy, the DOS departs from a smooth behavior.
For three-dimensional systems, 
such departure has a typical square-root energy dependence. 
For two-dimensional systems, there are two typical types of Van Hove singularity, one for which the departure of the DOS from a smooth behavior has a step, while for the other such departure has a logarithmic divergence. 
For one-dimensional systems the departure typically behaves as the inverse square-root at the Van Hove singularity energy.

In the section SuppMat3, we treat the case of the Sommerfeld expansion when the Fermi energy is precisely at a Van Hove singularity. 
The expansion of the integral
Eq.(\ref{eq:sommerfeld-integral-up-to-2})
for the 3D and 1D cases, as well as for the 2D-step case, is modified as follows
\begin{eqnarray}
I
&=&
\int_{-\infty}^{\mu} h(\varepsilon) d\varepsilon + C_{h,D}(\mu)(k_\textrm{B}T)^{D/2} + {\cal{O}}(k_\textrm{B}T)^{D/2+\delta^+},
\nonumber
\\
\label{eq:sommerfeld-integral-VanHove}
\end{eqnarray}
where $\delta^+$, a strictly positive number, and $C_{h,D}(\mu)$ depends on the function $h$ and the dimensionality $D$ of the system. $C_{h,D}(\mu)$ is given in the section SuppMat3. 

Thus, if the DOS is smooth, independently of the 
dimensionality, the lowest temperature dependence is proportional to $(k_\textrm{B}T)^{2}$. If the Fermi energy is at a 3D square-root singularity, the lowest temperature dependence is
proportional to $(k_\textrm{B}T)^{3/2}$, while for a 2D-step DOS, this become proportional to $k_\textrm{B}T$, and finally to $(k_\textrm{B}T)^{1/2}$ for a 1D inverse-square-root singularity.

\section{Low-temperature density-functional theory for metals based on the Sommerfeld expansion}
\label{sec:lowT-DFT}

We now consider the low-temperature expansion of the density, Hamiltonian,
eigenenergies, ...
for periodic metals, 
for vanishing reference temperature $T^{(0)}=0K$, within DFT, thus addressing the consequences of self-consistency. We first suppose that the DOS is smooth at the Fermi energy, avoiding the treatment of systems where this Fermi energy is at a Van Hove singularity at zero Kelvin. 
Modifications will follow straighforwardly for the Van Hove singularity case.

We write generically
\begin{eqnarray}
X(T) = X^{T=0}+T^{\gamma} X^{(\Delta T)}
+
\mathcal{O}(T^{\gamma+\delta^+}),
\label{eq:Xgeneric}
\end{eqnarray}
with vanishing $X^{(\Delta T)}$ at $T=0$,
and examine the first-order deviations in $T^{\gamma}$, denoted with a $(\Delta T)$ superscript,
$X^{(\Delta T)}$, of the different quantities, from 
their value at $T=0$.
Additional variations with a higher $T$-power than $\gamma$ will be neglected. 
The methodology to obtain these results is the same as in Sec.\ref{sec:DFPT}, although the presence of the
$\gamma$ exponent to characterize the
lowest-order $T$-dependence is a departure from usual DFPT expressions.
From the different equations of DFT
combined with the Sommerfeld expansion, in what follows,
we obtain 
that the lowest-order $X^{(\Delta T)}$ is quadratic in $T$, hence $\gamma=2$, if the DOS is smooth.

As mentioned at the end of Sec.\ref{sec:DFPT},
the inverse transpose of the electronic
dielectric operator governs the self-consistency. It does not present 
a singular behavior 
in the limit $T^{(0)}\to 0$.
The presence of $1/T^{(0)}$ in the first term of the rightmost factor in Eq.(\ref{eq:147})
(also present in (also present in Eqs.(\ref{eq:177}),(\ref{eq:mu1}) and (\ref{eq:F2mixed})) 
triggers a more serious challenge, and is addressed thanks to the
Sommerfeld expansion.
This is done in the section SuppMat5, including the self-consistency behavior.

One obtains the following expression for the
bare change of density due to a temperature change:
\begin{align}
\Delta\rho_{\textrm{bare}}(\textbf{r}) =& 
\frac{\pi^2}{6}
(k_\textrm{B}T)^2 
\nonumber\\
\times &
\Bigg(
\frac{
\partial \rho^{T=0}(\varepsilon,\textbf{r})}
{\partial \varepsilon}
\Big|_{\varepsilon_{\textrm{F}}}
-
\frac{
\rho^{T=0}(\varepsilon_{\textrm{F}},\textbf{r})
}
{g_{\textrm{DOS}}(\varepsilon_{\textrm{F}})}
\frac{
\partial g_{\textrm{DOS}}}
{\partial \varepsilon}
\Big|_{\varepsilon_{\textrm{F}}}
\Bigg).
\label{eq:rhoNSC}
\end{align}

This expression shows that the density change comes from the energy derivative of the l-DOS at the Fermi energy, albeit with a correction needed to insure charge neutrality of this l-DOS density change. 
Indeed, the second term in parentheses in Eq.(\ref{eq:rhoNSC}) is such that the integral of 
$\rho^{(\Delta T)}_{\textrm{bare}}(\textbf{r})$ over the whole space vanishes, as
\begin{eqnarray} 
\int
\frac{
\partial \rho^{T=0}(\textbf{r}',\varepsilon)}
{\partial \varepsilon}
\Big|_{\varepsilon_{\textrm{F}}}
d\textbf{r}'
=
\int
\frac{
\rho^{T=0}(\textbf{r}',\varepsilon_{\textrm{F}})
}
{g_{\textrm{DOS}}(\varepsilon_{\textrm{F}})}
\frac{
\partial g_{\textrm{DOS}}}
{\partial \varepsilon}
\Big|_{\varepsilon_{\textrm{F}}}
d\textbf{r}',
\nonumber\\
\label{eq:ChNeut}
\end{eqnarray}
using Eq.(\ref{eq:gDOS_e_inhom_1}).
The expression of the second term of the expansion 
Eq.(\ref{eq:Xgeneric}) for the change of density, including the self-consistency effect, is 
(with $\gamma$=2)
\begin{align}
\rho^{(\Delta T)}(\textbf{r}) &=
\frac{\pi^2}{6}
k_\textrm{B}^2 
\int
\epsilon_{\textrm{eFth}}^{-1t}(\textbf{r},\textbf{r}')
\nonumber\\
\times& \Bigg(
\frac{
\partial \rho^{T=0}(\varepsilon,\textbf{r}')}
{\partial \varepsilon}
\Big|_{\varepsilon_{\textrm{F}}}
-
\frac{
\rho^{T=0}(\varepsilon_{\textrm{F}},\textbf{r}')
}
{g_{\textrm{DOS}}(\varepsilon_{\textrm{F}})}
\frac{
\partial g_{\textrm{DOS}}}
{\partial \varepsilon}
\Big|_{\varepsilon_{\textrm{F}}}
\Bigg)
d\textbf{r}'.
\label{eq:finalrho}
\end{align}

In the case of a Van Hove singularity,
Eq.(\ref{eq:sommerfeld-integral-VanHove}) delivers
\begin{align}
\Delta\rho_{\textrm{bare}}(\textbf{r}) =& 
C_{h,D}(\mu)(k_\textrm{B}T)^{D/2}
\nonumber\\
\times &
\Bigg(
\frac{
\partial \rho^{T=0}(\varepsilon,\textbf{r})}
{\partial \varepsilon}
\Big|_{\varepsilon_{\textrm{F}}}
-
\frac{
\rho^{T=0}(\varepsilon_{\textrm{F}},\textbf{r})
}
{g_{\textrm{DOS}}(\varepsilon_{\textrm{F}})}
\frac{
\partial g_{\textrm{DOS}}}
{\partial \varepsilon}
\Big|_{\varepsilon_{\textrm{F}}}
\Bigg).
\label{eq:rhoNSC_vanHove}
\end{align}


\section{Conclusion}
\label{Sec:Conclusion}

In the present work, the response of a metal to a temperature change has been examined thanks to 
density-functional perturbation theory and the Sommerfeld expansion.

When the reference temperature does not vanish, 
$T^{(0)} \neq 0 K$, 
the DFPT treatment of such temperature change
is quite similar to the treatment
of other perturbations.
A bare change of occupation numbers induces the modification
of the charge density. It is screened by the self-consistent 
response of the metal,
that includes the usual Adler-Wiser non-interacting susceptibility as well as the modification of the density brought by the Fermi level change. 
The presence of both Adler-Wiser and
Fermi level susceptibility had already been noted in other recent works on
DFT self-consistency for metals.
It is found that the change of many properties 
(density, eigenenergies, Hamiltonian, DOS, e.g.) is linear with temperature. This DFPT formalism at finite temperatures is implemented in DFTK, and validated.

By contrast, at $T^{(0)} = 0 K$, the situation is less clear, 
as several formulas in the DFPT formalism diverge as the inverse of the reference temperature. This is addressed thanks to the Sommerfeld expansion, that we also consider in the case where the DOS is not smooth, unlike in the original work of Sommerfeld.

We combine the Sommerfeld expansion with perturbation theory as well as with density-functional perturbation theory.
If the DOS is smooth at the Fermi energy, the bare density change and the self-consistent density change due to temperature at low temperatures have a $T$-quadratic dependence. When the Fermi energy is at a Van Hove singularity, another power law is found: 
if the Fermi energy is at a 3D square-root Van Hove singularity, the lowest temperature dependence is
proportional to $(k_\textrm{B}T)^{3/2}$, for a 2D-step DOS, this becomes proportional to $k_\textrm{B}T$, and finally to $(k_\textrm{B}T)^{1/2}$ for a 1D inverse-square-root singularity.


\begin{acknowledgments} 
This work has been supported by the Fonds de la Recherche
Scientifique (FRS-FNRS Belgium) through the PdR Grant No.
T.0103.19 – ALPS. 
It is an outcome of the Shapeable
2D magnetoelectronics by design project (SHAPEme, EOS
Project No. 560400077525) that has received funding from
the FWO and FRS-FNRS under the Belgian Excellence of
Science (EOS) program.
\end{acknowledgments}

\bibliography{main} 

@article{Wiser1963,
    author = {Nathan Wiser},
    journal = {Phys. Rev.},
    pages = {62},
    title = {Dielectric Constant with Local Field Effects Included},
    volume = {129},
    year = {1963},
    doi = {https://doi.org/10.1103/PhysRev.129.62}
}

@article{Adler1962,
    author = {Stephen L. Adler},
    journal = {Phys. Rev.},
    pages = {413},
    title = {Quantum Theory of the Dielectric Constant in Real Solids},
    volume = {126},
    year = {1962},
    doi = {https://doi.org/10.1103/PhysRev.126.413}
}

@article{Gonze2024a,
  author  = {Gonze, Xavier and Rostami, Samare and Tantardini, Christian},
  title   = {Variational density functional perturbation theory for metals},
  journal = {Phys. Rev. B},
  volume  = {109},
  pages   = {014317},
  year    = {2024},
  doi     = {10.1103/PhysRevB.109.014317},
  url     = {https://link.aps.org/doi/10.1103/PhysRevB.109.014317}
}

@article{Methfessel1989,
  author  = {Methfessel, M. and Paxton, A. T.},
  title   = {{High-precision sampling for Brillouin-zone integration in metals}},
  journal = {Phys. Rev. B},
  volume  = {40},
  number  = {6},
  pages   = {3616--3621},
  year    = {1989},
  doi     = {10.1103/PhysRevB.40.3616},
  url     = {https://link.aps.org/doi/10.1103/PhysRevB.40.3616}
}

@article{DosSantos2023,
  author  = {dos Santos, Flaviano Jos{\'e} and Marzari, Nicola},
  title   = {Fermi energy determination for advanced smearing techniques},
  journal = {Phys. Rev. B},
  volume  = {107},
  number  = {19},
  pages   = {195122},
  year    = {2023},
  doi     = {10.1103/PhysRevB.107.195122},
  url     = {https://link.aps.org/doi/10.1103/PhysRevB.107.195122}
}

@article{Sommerfeld1928,
  author  = {Sommerfeld, A.},
  title   = {{Zur Elektronentheorie der Metalle auf Grund der Fermischen Statistik}},
  journal = {Zeitschrift f{\"u}r Physik},
  volume  = {47},
  pages   = {1--3},
  year    = {1928},
  doi     = {10.1007/BF01391052},
  url     = {https://link.springer.com/article/10.1007/BF01391052}
}

@article{Gonze_2020,
  title   = {{The Abinit project: Impact, environment and recent developments}},
  author  = {Gonze, X. and Amadon, B. and Antonius, G. and Arnardi, F. and Baguet, L. and Beuken, J.-M. and Bieder, J. and Bottin, F. and Bouchet, J. and Bousquet, E. and Brouwer, N. and Bruneval, F. and Brunin, G. and Cavignac, T. and Charraud, J.-B. and Chen, W. and C{\^o}t{\'e}, M. and Cottenier, S. and Denier, J. and Geneste, G. and Ghosez, P. and Giantomassi, M. and Gillet, Y. and Gingras, O. and Hamann, D. R. and Hautier, G. and He, X. and Helbig, N. and Holzwarth, N. and Jia, Y. and Jollet, F. and Lafargue-Dit-Hauret, W. and Lejaeghere, K. and Marques, M. A. L. and Martin, A. and Martins, C. and Miranda, H. P. C. and Naccarato, F. and Persson, K. and Petretto, G. and Planes, V. and Pouillon, Y. and Prokhorenko, S. and Ricci, F. and Rignanese, G.-M. and Romero, A. H. and Schmitt, M. M. and Torrent, M. and van Setten, M. J. and Van Troeye, B. and Verstraete, M. J. and Z{\'e}rah, G. and Zwanziger, J. W.},
  journal = {Computer Physics Communications},
  volume  = {248},
  pages   = {107042},
  year    = {2020},
  doi     = {10.1016/j.cpc.2019.107042},
  url     = {https://doi.org/10.1016/j.cpc.2019.107042},
}

@article{Romero2020,
  author  = {Romero, Aldo H. and Allan, Douglas C. and Amadon, Bernard and Antonius, Gabriel and Applencourt, Thomas and Baguet, Lucas and Bieder, Jordan and Bottin, Fran{\c c}ois and Bouchet, Johann and Bousquet, Eric and Bruneval, Fabien and Brunin, Guillaume and Caliste, Damien and C{\^o}t{\'e}, Michel and Denier, Jules and Dreyer, Cyrus and Ghosez, Philippe and Giantomassi, Matteo and Gillet, Yannick and Gingras, Olivier and Hamann, Donald R. and Hautier, Geoffroy and Jollet, Fran{\c c}ois and Jomard, G{\'e}rald and Martin, Alexandre and Miranda, Henrique P. C. and Naccarato, Francesco and Petretto, Guido and Pike, Nicholas A. and Planes, Valentin and Prokhorenko, Sergei and Rangel, Tonatiuh and Ricci, Fabio and Rignanese, Gian-Marco and Royo, Miquel and Stengel, Massimiliano and Torrent, Marc and van Setten, Michiel J. and Van Troeye, Benoit and Verstraete, Matthieu J. and Wiktor, Julia and Zwanziger, Josef W. and Gonze, Xavier},
  journal = {J. Chem. Phys.},
  pages   = {124102},
  title   = {{ABINIT: Overview, and focus on selected capabilities}},
  volume  = {152},
  year    = {2020},
  doi     = {10.1063/1.5144261},
  url     = {https://doi.org/10.1063/1.5144261},
}

@article{Verstraete2001,
  author  = {Verstraete, Matthieu and Gonze, Xavier},
  title   = {Smearing scheme for finite-temperature electronic-structure calculations},
  journal = {Phys. Rev. B},
  volume  = {65},
  pages   = {035111},
  year    = {2001},
  doi     = {10.1103/PhysRevB.65.035111},
  url     = {https://link.aps.org/doi/10.1103/PhysRevB.65.035111},
}

@article{Gonze1997a,
  author  = {Gonze, Xavier and Lee, Changyol},
  title   = {Dynamical matrices, Born effective charges, dielectric permittivity tensors, and interatomic force constants from density-functional perturbation theory},
  journal = {Phys. Rev. B},
  volume  = {55},
  number  = {16},
  pages   = {10355--10368},
  year    = {1997},
  doi     = {10.1103/PhysRevB.55.10355},
  url     = {https://link.aps.org/doi/10.1103/PhysRevB.55.10355},
}

@article{Gonze1997,
  author  = {Gonze, Xavier},
  title   = {First-principles responses of solids to atomic displacements and homogeneous electric fields: {Implementation} of a conjugate-gradient algorithm},
  journal = {Phys. Rev. B},
  volume  = {55},
  number  = {16},
  pages   = {10337--10354},
  year    = {1997},
  doi     = {10.1103/PhysRevB.55.10337},
  url     = {https://link.aps.org/doi/10.1103/PhysRevB.55.10337},
}

@article{Baroni2001,
  title   = {Phonons and related crystal properties from density-functional perturbation theory},
  author  = {Baroni, Stefano and de Gironcoli, Stefano and Dal Corso, Andrea and Giannozzi, Paolo},
  journal = {Rev. Mod. Phys.},
  volume  = {73},
  number  = {2},
  pages   = {515--562},
  year    = {2001},
  doi     = {10.1103/RevModPhys.73.515},
  url     = {https://link.aps.org/doi/10.1103/RevModPhys.73.515}
}

@article{Marzari1997,
  author  = {Marzari, Nicola and Vanderbilt, David and Payne, M. C.},
  title   = {Ensemble Density-Functional Theory for Ab Initio Molecular Dynamics of Metals and Finite-Temperature Insulators},
  journal = {Phys. Rev. Lett.},
  volume  = {79},
  pages   = {1337--1340},
  year    = {1997},
  doi     = {10.1103/PhysRevLett.79.1337},
  url     = {https://link.aps.org/doi/10.1103/PhysRevLett.79.1337}
}

@article{Feynman1939,
  author  = {Feynman, R. P.},
  title   = {Forces in Molecules},
  journal = {Phys. Rev.},
  volume  = {56},
  pages   = {340--343},
  year    = {1939},
  doi     = {10.1103/PhysRev.56.340},
  url     = {https://link.aps.org/doi/10.1103/PhysRev.56.340}
}

@book{Hellmann1937,
  title     = {Einf{\"u}hrung in die Quantenchemie},
  author    = {Hellmann, H.},
  year      = {1937},
  publisher = {Deuticke},
  address   = {Leipzig}
}

@article{Verstraete2004,
  title   = {Metals at finite temperature: a modified smearing scheme},
  author  = {Verstraete, Matthieu and Gonze, Xavier},
  journal = {Computational Materials Science},
  volume  = {30},
  number  = {1},
  pages   = {27--33},
  year    = {2004},
  doi     = {10.1016/j.commatsci.2004.01.006},
  url     = {https://doi.org/10.1016/j.commatsci.2004.01.006}
}

@article{mermin1965thermal,
  title   = {Thermal properties of the inhomogeneous electron gas},
  author  = {Mermin, N. David},
  journal = {Phys. Rev.},
  volume  = {137},
  number  = {5A},
  pages   = {A1441--A1443},
  year    = {1965},
  doi     = {10.1103/PhysRev.137.A1441},
  url     = {https://link.aps.org/doi/10.1103/PhysRev.137.A1441}
}

@article{Gonze1989_2nplus1,
  title   = {Density-functional approach to nonlinear-response coefficients of solids},
  author  = {Gonze, X. and Vigneron, J.-P.},
  journal = {Phys. Rev. B},
  volume  = {39},
  pages   = {13120--13128},
  year    = {1989},
  doi     = {10.1103/PhysRevB.39.13120},
  url     = {https://link.aps.org/doi/10.1103/PhysRevB.39.13120}
}

@article{Gonze1995_PRA1086,
  title   = {Perturbation expansion of variational principles at arbitrary order},
  author  = {Gonze, Xavier},
  journal = {Phys. Rev. A},
  volume  = {52},
  pages   = {1086--1095},
  year    = {1995},
  doi     = {10.1103/PhysRevA.52.1086},
  url     = {https://link.aps.org/doi/10.1103/PhysRevA.52.1086}
}

@article{Gonze1995_PRA1096,
  title   = {Adiabatic density-functional perturbation theory},
  author  = {Gonze, Xavier},
  journal = {Phys. Rev. A},
  volume  = {52},
  pages   = {1096--1114},
  year    = {1995},
  doi     = {10.1103/PhysRevA.52.1096},
  url     = {https://link.aps.org/doi/10.1103/PhysRevA.52.1096}
}

@article{schmitz2025algorithmic,
  title         = {{Algorithmic differentiation for plane-wave DFT: materials design, error control and learning model parameters}},
  author        = {Schmitz, Niklas Frederik and Ploumhans, Bruno and Herbst, Michael F.},
  journal       = {arXiv},
  year          = {2025},
  eprint        = {2509.07785},
  archivePrefix = {arXiv},
  url           = {https://arxiv.org/abs/2509.07785}
}

@inproceedings{herbst2021dftk,
  title   = {{DFTK: A Julian approach for simulating electrons in solids}},
  author  = {Herbst, Michael F. and Levitt, Antoine and Canc{\`e}s, Eric},
  booktitle = {Proceedings of the JuliaCon Conferences},
  volume  = {3},
  pages   = {69},
  year    = {2021},
  note    = {Issue 26},
  doi     = {10.21105/jcon.00069},
  url     = {https://doi.org/10.21105/jcon.00069}
}

@article{revels2016forward,
  title         = {{Forward-mode automatic differentiation in Julia}},
  author        = {Revels, Jarrett and Lubin, Miles and Papamarkou, Theodore},
  journal       = {arXiv},
  year          = {2016},
  eprint        = {1607.07892},
  archivePrefix = {arXiv},
  url           = {https://arxiv.org/abs/1607.07892}
}

@article{Tantardini2025a,
  author  = {Tantardini, Christian and Kvashnin, Alexander G. and Giantomassi, Matteo and Ilia{\v s}, Miroslav and Yakobson, Boris I. and Hemley, Russell J. and Gonze, Xavier},
  title   = {{Charge density waves and structural phase transition in the high-$T_c$ superconducting LaH$_{10}$ quantum crystal}},
  journal = {Phys. Rev. B},
  volume  = {112},
  pages   = {115154},
  year    = {2025},
  doi     = {10.1103/PhysRevB.112.115154},
  url     = {https://link.aps.org/doi/10.1103/PhysRevB.112.115154}
}

@book{Mahan2000,
  author    = {Mahan, Gerald D.},
  title     = {Many-Particle Physics},
  edition   = {3},
  year      = {2000},
  publisher = {Springer},
  address   = {Boston, MA},
  isbn      = {978-0-306-46380-8},
  doi       = {10.1007/978-1-4757-5714-9},
  url       = {https://doi.org/10.1007/978-1-4757-5714-9}
}

@book{FetterWalecka2018,
  author    = {Fetter, Alexander L. and Walecka, John Dirk},
  title     = {Quantum Theory of Many-Particle Systems},
  year      = {2018},
  publisher = {CRC Press},
  address   = {Boca Raton, FL},
  isbn      = {978-0-201-36035-1},
  doi       = {10.1201/9780429493218},
  url       = {https://doi.org/10.1201/9780429493218}
}

@book{Huang1987,
  author    = {Huang, Kerson},
  title     = {Statistical Mechanics},
  edition   = {2},
  year      = {1987},
  publisher = {John Wiley \& Sons},
  address   = {New York},
  isbn      = {978-0-471-81518-1},
  doi       = {10.1002/9783527617259},
  url       = {https://doi.org/10.1002/9783527617259}
}

@article{Gironcoli1995,
  author  = {de Gironcoli, S.},
  title   = {Lattice dynamics of metals from density-functional perturbation theory},
  journal = {Phys. Rev. B},
  volume  = {51},
  pages   = {6773--6776},
  year    = {1995},
  doi     = {10.1103/PhysRevB.51.6773},
  url     = {https://link.aps.org/doi/10.1103/PhysRevB.51.6773}
}

@article{Baroni1987,
  author  = {Baroni, S. and Giannozzi, P. and Testa, A.},
  title   = {Green's-Function Approach to Linear Response in Solids},
  journal = {Phys. Rev. Lett.},
  volume  = {58},
  pages   = {1861--1864},
  year    = {1987},
  doi     = {10.1103/PhysRevLett.58.1861},
  url     = {https://link.aps.org/doi/10.1103/PhysRevLett.58.1861}
}

@article{Giannozzi2017,
  author  = {Giannozzi, P. and Andreussi, O. and Brumme, T. and Bunau, O. and Buongiorno Nardelli, M. and Calandra, M. and Car, R. and Cavazzoni, C. and Ceresoli, D. and Cococcioni, M. and Colonna, N. and Carnimeo, I. and Dal Corso, A. and de Gironcoli, S. and Delugas, P. and DiStasio Jr., R. A. and Ferretti, A. and Floris, A. and Fratesi, G. and Fugallo, G. and Gebauer, R. and Gerstmann, U. and Giustino, F. and Gorni, T. and Jia, J. and Kawamura, M. and Ko, H.-Y. and Kokalj, A. and K{\"u}{\c c}{\"u}kbenli, E. and Lazzeri, M. and Marsili, M. and Marzari, N. and Mauri, F. and Nguyen, N. L. and Nguyen, H.-V. and Otero-de-la-Roza, A. and Paulatto, L. and Ponc{\'e}, S. and Rocca, D. and Sabatini, R. and Santra, B. and Schlipf, M. and Seitsonen, A. P. and Smogunov, A. and Timrov, I. and Thonhauser, T. and Umari, P. and Vast, N. and Wu, X. and Baroni, S.},
  title   = {Advanced capabilities for materials modelling with {Quantum ESPRESSO}},
  journal = {J. Phys.: Condens. Matter},
  volume  = {29},
  pages   = {465901},
  year    = {2017},
  doi     = {10.1088/1361-648X/aa8f79},
  url     = {https://doi.org/10.1088/1361-648X/aa8f79}
}

@article{Gonze2005a,
  author  = {Gonze, X. and Rignanese, G.-M. and Caracas, R.},
  title   = {First-principle studies of the lattice dynamics of crystals, and related properties},
  journal = {Zeitschrift f{\"u}r Kristallographie},
  volume  = {220},
  number  = {5-6},
  pages   = {458--472},
  year    = {2005},
  doi     = {10.1524/zkri.220.5.458.65077},
  url     = {https://doi.org/10.1524/zkri.220.5.458.65077}
}

@article{cances2023numerical,
  title   = {Numerical stability and efficiency of response property calculations in density functional theory},
  author  = {Canc{\`e}s, Eric and Herbst, Michael F. and Kemlin, Gaspard and Levitt, Antoine and Stamm, Benjamin},
  journal = {Letters in Mathematical Physics},
  volume  = {113},
  number  = {1},
  pages   = {21},
  year    = {2023},
  doi     = {10.1007/s11005-023-01645-3},
  url     = {https://doi.org/10.1007/s11005-023-01645-3}
}

@article{gerard1990resonance,
  title   = {{Resonance theory for periodic Schr{\"o}dinger operators}},
  author  = {G{\'e}rard, Christian},
  journal = {Bulletin de la Soci{\'e}t{\'e} math{\'e}matique de France},
  volume  = {118},
  number  = {1},
  pages   = {27--54},
  year    = {1990},
  url     = {https://www.numdam.org/item/BSMF_1990__118_1_27_0/}
}

@book{ashcroft1976,
  author    = {Ashcroft, Neil W. and Mermin, N. David},
  title     = {Solid State Physics},
  year      = {1976},
  publisher = {Holt, Rinehart and Winston},
  address   = {New York},
  isbn      = {978-0-03-083993-1},
  note      = {Also appears in later unaltered reprints; other ISBNs include 0-03-083993-9 and 0-03-049346-3.},
  url       = {https://katalog.ub.uni-heidelberg.de/titel/65976347}
}

@book{inkson1986,
  author    = {Inkson, John C.},
  title     = {Many-Body Theory of Solids: An Introduction},
  publisher = {Springer},
  address   = {Boston, MA},
  year      = {1984},
  doi       = {10.1007/978-1-4757-0226-2},
  isbn      = {978-1-4757-0226-2},
  url       = {https://link.springer.com/book/10.1007/978-1-4757-0226-2}
}

@book{Martin2004,
  author    = {Martin, Richard M.},
  title     = {Electronic Structure: Basic Theory and Practical Methods},
  publisher = {Cambridge University Press},
  address   = {Cambridge},
  year      = {2004},
  isbn      = {978-0-521-78285-2}
}

@article{Zabalo2024a,
  author  = {Zabalo, Asier and Stengel, Massimiliano},
  title   = {Ensemble density functional perturbation theory: Spatial dispersion in metals},
  journal = {Phys. Rev. B},
  volume  = {109},
  pages   = {245116},
  year    = {2024},
  doi     = {10.1103/PhysRevB.109.245116},
  url     = {https://link.aps.org/doi/10.1103/PhysRevB.109.245116}
}

@article{Herbst2020a,
  author  = {Herbst, Michael F. and Levitt, Antoine},
  title   = {Black-box inhomogeneous preconditioning for self-consistent field iterations in density functional theory},
  journal = {J. Phys.: Condens. Matter},
  volume  = {33},
  pages   = {085503},
  year    = {2021},
  doi     = {10.1088/1361-648X/abcbdb},
  url     = {https://doi.org/10.1088/1361-648X/abcbdb}
}

@article{VanHove1953,
  author  = {Van Hove, L{\'e}on},
  title   = {The Occurrence of Singularities in the Elastic Frequency Distribution of a Crystal},
  journal = {Phys. Rev.},
  volume  = {89},
  pages   = {1189},
  year    = {1953},
  doi     = {10.1103/PhysRev.89.1189},
  url     = {https://link.aps.org/doi/10.1103/PhysRev.89.1189}
}

@book{Parr1989,
  author    = {Parr, Robert G. and Yang, Weitao},
  title     = {Density-Functional Theory of Atoms and Molecules},
  publisher = {Oxford University Press},
  address   = {New York},
  year      = {1989},
  isbn      = {0-19-504279-4}
}

@article{Gonis2018,
  author  = {Gonis, A. and D{\"a}ne, M.},
  title   = {{Extension of the Kohn-Sham formulation of density functional theory to finite temperature}},
  journal = {J. Phys. Chem. Solids},
  volume  = {116},
  pages   = {86--99},
  year    = {2018},
  doi     = {10.1016/j.jpcs.2017.12.021}
}

@article{Perrot1984,
  author  = {Perrot, F. and Dharma-wardana, M. W. C.},
  title   = {Exchange and correlation potentials for electron-ion systems at finite temperatures},
  journal = {Phys. Rev. A},
  volume  = {30},
  pages   = {2619},
  year    = {1984},
  doi     = {10.1103/PhysRevA.30.2619},
  url     = {https://link.aps.org/doi/10.1103/PhysRevA.30.2619}
}

@article{Brown2013,
  author  = {Brown, Ethan W. and Clark, Bryan K. and DuBois, Jonathan L. and Ceperley, David M.},
  title   = {Path-Integral Monte Carlo Simulation of the Warm Dense Homogeneous Electron Gas},
  journal = {Phys. Rev. Lett.},
  volume  = {110},
  pages   = {146405},
  year    = {2013},
  doi     = {10.1103/PhysRevLett.110.146405},
  url     = {https://link.aps.org/doi/10.1103/PhysRevLett.110.146405}
}

@article{Karasiev2014,
  author  = {Karasiev, Valentin V. and Sjostrom, Travis and Dufty, James and Trickey, S. B.},
  title   = {Accurate Homogeneous Electron Gas Exchange-Correlation Free Energy for Local Spin-Density Calculations},
  journal = {Phys. Rev. Lett.},
  volume  = {112},
  pages   = {076403},
  year    = {2014},
  doi     = {10.1103/PhysRevLett.112.076403},
  url     = {https://link.aps.org/doi/10.1103/PhysRevLett.112.076403}
}

@article{Karasiev2018,
  author  = {Karasiev, Valentin V. and Dufty, James W. and Trickey, S. B.},
  title   = {Nonempirical Semilocal Free-Energy Density Functional for Matter under Extreme Conditions},
  journal = {Phys. Rev. Lett.},
  volume  = {120},
  pages   = {076401},
  year    = {2018},
  doi     = {10.1103/PhysRevLett.120.076401}
}

\end{document}